\documentclass[aps,preprint]{revtex4}%
\usepackage{amsfonts}
\usepackage{amsmath}
\usepackage{amssymb}
\usepackage{graphicx}%
\providecommand{\U}[1]{\protect\rule{.1in}{.1in}}

\begin{document}
\preprint{ }
\title[ ]{Entanglement and  Minimal Hilbert Space in  the Classical Dual States of Quantum Theory}

\author{Diego J. CIRILO-LOMBARDO}
\affiliation{Special Astrophysical Observatory of Russian Academy of Sciences,\\
Radio Astrophysics Laboratory, AGN research group, Nizhny Arkhyz, Russian
Federation}
\affiliation{M. V. Keldysh Institute of the Russian Academy of Sciences, Federal Research
Center-Institute of Applied Mathematics, Miusskaya sq. 4, 125047 Moscow,
Russian Federation}
\affiliation{CONICET-Universidad de Buenos Aires, Departamento de Fisica, Instituto de
Fisica Interdisciplinaria y Aplicada (INFINA), Buenos Aires, Argentina.}
\author{Norma G. SANCHEZ}
\affiliation{The International School of Astrophysics Daniel Chalonge - Hector de Vega, CNRS,
INSU-Institut National des Sciences de l'Univers,  Sorbonne University.  75014
Paris, France.\\
Norma.Sanchez@obspm.fr \\
https://chalonge-devega.fr/sanchez
}
\;
\date{\today}
\pacs{PACS number}

\begin{abstract} 
A precise physical description and understanding of {\bf the classical dual content of quantum theory} is necessary in many disciplines today: from  concepts and interpretation to quantum technologies and computation. In this paper we investigate Quantum Entanglement with the new approach APL Quantum 2, 016104 (2025) \url{https://doi.org/10.1063/5.0247698} on dual Classicalization. Thus, the results of this paper are twofold:  {\bf Entanglement and 
 Classicalization and the relationship between them}. Classicalization truly occurs {\it only} under the action of the Metaplectic group $Mp(n)$ (Minimal Representation group, double covering of the Symplectic group). {\bf The results of this paper are}: {\bf(1)} We compute and analyze Entanglement for different types of coherent (coset and non 
 coset) states and topologies: in the circle and the cylinder.  We project 
 the entangled wave functions onto the even (+) and odd (-) irreducible Hilbert $Mp(n)$ subspaces, and compute their  square norms: {\bf Entanglement Probabilities}   $P_{++}$,   $P_{--}$, $P_{+-}$, (eg in the same or in the different subspaces), and the Total sum of them.
{\bf (2) Entanglements in the circle}:  (i) of orthogonal states are $\neq  0$ even for the control phase $\rho = 0$ and are preserved for any phase $\rho \neq \pi /2$. $P_{+-}$ is separable (entanglement is broken) only if $\rho = 0$. (ii) For coincident states, entanglements are $\neq 0$ and separable for any control parameter $\rho \ne 0$. If  $\rho = 0 $, all three  $P_{++}$, $P_{--}$, $P_{+-}$ vanish.  (iii) {\it The Entanglements of coset circle states} depend on the angles and the coherent complex displacement parameter $\alpha$ through a {\it single variable} $z^{\prime}$ which condition on the disk $\left\vert z^{\prime}\right\vert < 1$ guarantees {\it both analyticity and normalization}. (iv) {\bf Entanglement classicalization} in the circle is {\bf stronger} for coset states than for non coset ones: a tail decreasing with increasing $n$ is absent in the non coset case. {\bf (3) Entanglement in the cylinder} depends weakly on the angles  ($\varphi -\varphi ^{\prime }$) and {\bf strongly classicalizes}: rapid exponential decay suppresion $e^{-n^{2}}$ for large $n$, stronger than in the circle (be coset or not). 
It expresses in terms of {\bf Theta functions} $\vartheta_{2}\left( 0,e^{-8}\right)$\, and \, $\vartheta_{3}\left( 0,e^{-8}\right)$\, in {\bf all} classicalization projections in the degeneracy (equal states) limit. 
{\bf (4)} Comparison with other (non $Mp(2)$) Entanglements, as the Schrodinger cat states, show very clear differences, (eg. Figs \ref{f3}  and \ref{f4}).  {\bf (5)} The {\bf Antipodal and Non Antipodal Entanglements} (as regulated by the phase control parameter $\rho = \pi$ and $\rho = 0$ respectively), are clearly different in all types (circle, cylinder, coset or not) of {\bf orthogonal} states studied here. {\it The Antipodal  Entanglement is the Minimal}. These theoretical and conceptual results can be of experimental and practical real-world interest. 
\end{abstract}
\volumeyear{year}
\volumenumber{number}
\issuenumber{number}
\eid{identifier}
\date[Date ]{august 2025}

\maketitle
\tableofcontents

\section{Introduction and Results}

 Quantum theory is at the center of science and technology today:  The great success of Quantum theory and its deep and wide impact transcends physics and its applications: See for example Refs \cite{Hess-2024}, \cite{Nobel-2022}, \cite{Nobel-2024}, \cite {MacLoughlin-Sanchez}  \cite{IYQ2025}. While the quantum wave nature of particles and the intrinsic quantum uncertainty principle were mainly about the first quantum revolution, the superposition principle and quantum entanglement are at the heart of the second quantum revolution at work.
While quantum research is rapidly evolving in many different directions and disciplines, it is necessary to have a precise modern physical understanding and clear computational framework for {\bf the classical dual content of quantum theory}: This impacts from the conceptual and quantum interpretation measurements until quantum technologies and computation research. Therefore, one fundamental question in Quantum theory with interest for its meaning, observation and experimental research is the following:

\begin{center}
Under which conditions Quantum theory does {\bf appear Classical, namely which are the Classical dual sectors of the Quantum world}.
\end{center}
Classical states are a particular case of Quantum theory and are dual   states in the precise sense of the general classical-quantum duality of Nature. Recently, in ApL Quantum 2025 Ref \cite{APL25}, we provided a precise answer to this problem within a novel  approach: The minimal group representation principle, which is uniquely realized by the Metaplectic group $Mp(2d)$. This group is the double covering of the Symplectic group $ Sp(2d)$ and its action 
inmediately {\it classicalizes} the system. We performed an extensive study with different types of quantum states on the circle and on the cylinder from which emerged too that $Mp(2d)$ is the symmetry group of the general classical-quantum duality of Nature.   

\bigskip

{\bf In this paper} we go beyond in our study of the Classical sectors of Quantum theory by investigating the {\bf Entanglement} for the quantum-classical dual states with topologies on the circle and on the cylinder and different types of coherent (coset and non coset) states. Thus, the results of
this paper are on both:
\begin{center}
{\bf Entanglement and Classicalization, and the  relationship between them}.
\end{center} 
This is important too for the bridge  between the classical and quantum information processings and relates to a real-world problem.

\medskip

{\bf Discretization} arises naturally and directly from the basic states of the metaplectic representations: the decomposition of the $Mp(2)$ group into its two irreducible representations span both: the \textit{even} $\left\vert
\;2n\;\right\rangle $ and \textit{odd} $\left\vert \;2n\;+\;1\;\right\rangle $ 
states respectively, ($n = 1,\,2,\,3\,...$)  of the harmonic oscillator, totally covered by the
metaplectic group. The two $Mp(2)$ irreducible subspaces contain both: the quantum and the classical dual sectors. For $n \rightarrow \infty$, the spectrum becomes naturally continuum as it must be.

\medskip

A quantum state  {\bf 
 completely  classicalizes}  its inherent quantum structure {\bf only} under the action of the {\bf Mp(n) (metaplectic)   Group}, (Minimal Representation Group).
Classicalization is explicit in the decreasing exponential factors for large $n$ arising in the $Mp(2)$ projections of the states: screenings
$e^{-2\,n}$, $e^{-(\,2n+1/2\,)}$   or $e^{-2\,n^{2}}$, $e^{-(\,2n+1/2\,)^{2}}$   (depending on the topology, circle or cylinder states respectively). The Classical-Quantum Duality is realized in the $ Mp(n)$ symmetry because of the complete covering of the Hilbert space: 
Each of the two, {\it even} $\mathcal{H}_{(+)}$ and {\it odd} $\mathcal{H}_{(-)}$ sectors are local coverings,  their sum being  global, completely covering the {\it whole} Hilbert space $\mathcal{H}$ = $\mathcal{H}_{(+)}%
\oplus\mathcal{H}_{(-)}$.  
  The two $(+)$ and $(-)$  sectors are classical-quantum duals of each other and are {\bf entangled}. This is also important in order to include  gravity at the Planck scale and beyond: quantum space-time and classical-quantum gravity duality, Refs \cite{NSPRD2021}, \cite{NSPRD2023}, \cite{Sanchez2019a}, \cite{Sanchez2019b}, \cite{Sanchez2019c} which is general, irrespective of the number or type of space-time dimensions or manifolds  (with or without compactifications).
\begin{center}
{\bf In this paper our focus and results are the following:}
\end{center}
\bigskip

{\bf(1) In analogy} and for comparison with the entanglement of the standard coherent
states in the number representation: 
$$\left\vert n,m\right\rangle
\;=\;\frac{1}{\sqrt{2}}\left(\, \left\vert n\right\rangle \otimes\left\vert
m\right\rangle +\left\vert m\right\rangle \otimes\left\vert n\right\rangle
\,\right),  $$
we project the states (be coherent or not)
into the number representation but through the basic states of the Metaplectic
group $Mp(n)$, namely:

\begin{equation}\label{Metastate}
\left\vert \Psi^{(\pm)}\left(  \omega\right)  \Psi^{(\pm
)}\left(  \sigma\right)  \right\rangle\; =\;\frac{1}{\sqrt{2}}\left( \, \left\vert
\Psi^{(\pm)}\left(  \omega\right)  \right\rangle \otimes\left\vert \Psi
^{(\pm)}\left(  \sigma\right)  \right\rangle +\left\vert \Psi^{(\pm)}\left(
\sigma\right)  \right\rangle \otimes\left\vert \Psi^{(\pm)}\left(
\omega\right)  \right\rangle \,\right)  \end{equation} 

\bigskip

that is to say, with the application of the Minimal Group Representation which is precisely given by the Metaplectic group $Mp(n)$.

\bigskip

{\bf(2) The entanglement} of the item (1) with the basic even (+) and odd (-) states of $Mp(n)$
 is compared through the degree of purity with
the case of the (even and odd) Schrodinger cat states: \begin{equation*}
\left\vert \alpha,-\alpha\right\rangle \;=\;\frac{1}{\sqrt{2}}\,\left(\,
\left\vert \alpha\right\rangle \otimes\left\vert -\alpha\right\rangle
\,+\,\left\vert -\alpha\right\rangle \otimes\left\vert \alpha\right\rangle
\,\right) .
\end{equation*}

 \bigskip

\textbf{(3) The entanglement} of two states with the topology of the circle
(or any other state, coherent or not) when projected according to the
Minimal Group Representation (Metaplectic group) provides two possibilities:
This is because from the point of view of the metaplectic projection we have
a general case entailing two analytical functions $(\sigma ,\omega)$ $%
\rightarrow\Psi\left( \omega\right) \Psi\left( \sigma\right)$, that is:
\begin{equation*}
\Psi\left( \omega\right) \left\{ 
\begin{array}{ccc}
& \Psi^{(+)}\left( \omega\right) \rightarrow & \mathcal{H}_{+} \\ 
&  &  \\ 
& \Psi^{(-)}\left( \omega\right) \longrightarrow & \mathcal{H}_{-}%
\end{array}
\right.
\;\;\;\;and\;\;\;\;
\Psi\left( \sigma\right) \left\{ 
\begin{array}{ccc}
& \Psi^{(+)}\left( \sigma\right) \rightarrow & \mathcal{H}_{+} \\ 
&  &  \\ 
& \Psi^{(-)}\left( \sigma\right) \longrightarrow & \mathcal{H}_{-}%
\end{array}
\right.
\end{equation*}
\newline
which gives precisely, (with the $Mp(n)$ schematic structure given
above), states of the type of Eq.(\ref{Metastate}). 

As a consequence of a characteristic property of the $Mp(n)$ group, this is the \textbf{only case} where {\bf a single
analytic function} allows to preserve the {\bf uniqueness} of the mapping with a
single analytic function, e.g. for
$\sigma\longrightarrow\omega$, then $\Psi\left(
\sigma\right)  \rightarrow\Psi\left(  \omega\right)  $. Therefore:\\
\begin{equation*}
\underset{\sigma\,\rightarrow \,\omega}{\lim}\,\left\vert \,\Psi^{(+)}\left(
\omega\right)  \Psi^{(-)}\left(  \sigma\right) \, \right\rangle \sim\,\frac
{1}{\sqrt{2}}\,\left[ \; \left\vert \Psi^{(+)}\left(  \omega\,\right)  \right\rangle
\otimes\left\vert \,\Psi^{(-)}\left(  \omega\,\right)  \right\rangle \,+\,\left\vert
\Psi^{(+)}\left(  \omega\right)  \right\rangle \otimes\left\vert \Psi
^{(-)}\left(  \omega\right)  \right\rangle \;\right]
\end{equation*} \\
This limit procedure must be taken explicitly in the projection for the
cases of the states on the cylinder, as we will see here in Section III.

\bigskip

{\bf (4) We compute and analyze} the different projections in each $Mp(2)$
even (+) and odd (-) sector of the entangled wave functions with topologies in the circle and in the cylinder. This provides for each topology three different entanglement possibilities depending on whether states are entangled in the same subspaces or in the crossed ones, namely  (++), (-\,-), or (+\,-), and the Total Entanglement of them.  We compute too the 
corresponding square norm {\bf Entanglement Probabilities} : even-even  $P_{++}$,  odd-odd  $P_{--}$, and crossed $P_{+-}$ Probabilities:

\bigskip

{\bf (5) For the circle (London)  states :} 
 The Entanglements $P_{+ +}$ or $P_{--}$ of the same subspace states are similar  (with $cosh$ and $cos$ functions for $P_{+ +}$ and  $sinh, sin$ functions for $P_{--}$, while the  crossed  $P_{+-}$ have both types of functions as it must be). We analyze the limits of coincident states  $\Delta \rightarrow 0$, and of orthogonal states $\Delta \rightarrow \pi /2$, being $\Delta \equiv \left( \varphi -\varphi ^{\prime }\right)$, and   the role of the phase  $\rho$ (the admissible control parameter which is useful for the entanglement experimental realizations).
 
{\bf For the coincident states $(\Delta\rightarrow \,0)$:} The Entanglements $P_{++}$, $P_{--}$, $P_{+\,-}$  are $\neq 0$ and  separable  (the entanglement is broken) for any phase  $\rho \ne 0$. While, all three  
$P_{++}$, $P_{--}$, $P_{+-}  = 0$  if  $\rho = 0 $.
{\bf For the orthogonal states $(\Delta \;\rightarrow \;\pi/2)$ :} The Entanglements $P_{++}$, $P_{--}$, $P_{+-}$ are $\neq  0$ even if  $\rho = 0$. $P_{++}$, $P_{--}$, $P_{+-}$ are preserved for any $\rho \neq \pi /2$. $P_{+-}$ is separable (the entanglement is broken) only if $\rho = 0$.  {\bf Figures \ref{f1} and \ref{f2}} and their captions illustrate some of these results.

\bigskip
 
{\bf (6) The Entanglements of the new coset circle states} $\left\vert \,\varphi, \alpha \right\rangle $ do depend on the angles $(\varphi, \varphi^{\prime} )$  and on the complex coherent parameter $\alpha $ through the {\bf single variable} $z^{\prime}\,=\,\omega\,e^{\,i\,\left(  \,\varphi
-\alpha^{\ast}/2\,\right) }$, which condition 
 on the disk $\left\vert z^{\prime}\right\vert \;<\;1$ does  remarkably guarantee both {\bf analyticity and  normalization}. This Entanglement clearly contains a decreasing tail as $n$ increases showing that {\bf the Entanglement classicalization} is {\bf stronger} in the coset states than in the non coset (London) states, which we relate to the fact that these coset circle states are truly coherent analytic and normalizable states, (while the London, 't Hooft states are not).

 \bigskip
 
 {\bf (7) The Entanglement of the cylinder states:} {\bf (i)}   Rapid  decay   through exponential suppresion factors $\varpropto $ $%
e^{-n^{2}}$ for large $n$, showing that in all cases the Cylinder Entanglement {\bf strongly classicalizes}. {\bf (ii)} The low dependence of the Entanglement on the angular variables (contained in $\Delta = (\varphi -\varphi ^{\prime })$) in the limiting case of degeneracy (equality) of the analytical functions describing the cylinder states : $P_{++}$ and $P_{--}$ are totally {\bf independent of the angles} $(\varphi ,\varphi ^{\prime })$, while the crossed $P_{+-}$ entanglement does depend weakly on them.  {\bf (iii)}   The cylinder Entanglements express in terms of {\bf Theta functions} $\vartheta_{2}\left( 0,e^{-8}\right)$\, and \, $\vartheta_{3}\left( 0,e^{-8}\right)$\, in {\bf all} classicalization projections in the degeneracy (equal states) limit. {\bf(iv)}   
Cylinder topology does classicalize the Entanglement {\bf stronger} than the circle topology. In {\bf Figure 1} we {\it shortly summarize some of the main Entanglement and  Classicalization features of this paper} both for the circle and cylinder topologies. 

\bigskip

{\bf (8) Comparison of Entanglements} are performed for completeness: For instance, $Mp(2)$ Entanglement projections are compared with other (non $Mp(2)$) projection states as  the Schrodinger cat states: Differences are very clear as shown in {\bf Figures \ref{f3} and \ref{f4}} and their captions.  The {\bf Antipodal and and Non Antipodal Entanglements}, (as regulated by the phase control parameter $\rho = \pi$ and $\rho = 0$ respectively),  are clearly different in the case of  orthogonal states for {\bf all types of orthogonal states} studied here 
 (being Schrodinger cat or Mp(2) orthogonal states).
 
\bigskip

{\bf(9) Besides} the theoretical and conceptual interest, the results of this paper should have {\bf applications and implications} for experimental entanglement work, connections of classical and quantum treatments of information, interpretation of the quantum-classical interaction, the quantum interpretation measurements, classical or quantum optimization, to mention some of them. In Section VI Concluding Remarks we mention some {\bf outlook} and implications of our results.

\bigskip

{\bf This paper is organized as follows:}
In Section II we describe and compute the quantum Entanglements of the phase states in the circle, its full global covering, the  Entanglement Probabilities and their classicalization. 
In Section III we fully compute and analyze the Entanglements of the Cylinder States, their Probabilities and show how their classicalization occurs, finding very different and new properties  with respect to the Entanglement and its classicalization in the phase space of the circle. In Section IV we perform the complete Entanglement for the general coset coherent states in the circle which allow to fully see the Entanglement and classicalizations conditions with respect to topology, analyticity in the disk of the wave functions and normalization, and compare with respect to the London (circle, phase space) states. In Section V we compare the Entanglements and Classicalization we found here with those of the Schrodinger cat states and discusses the implications of the $Mp(2)$ results of the previous sections with the cat state results. Sections VI  summarizes Remarks and Conclusions.
\begin{center}
\begin{figure}
\includegraphics[
height=9.7742in,
width=7.2029in,]{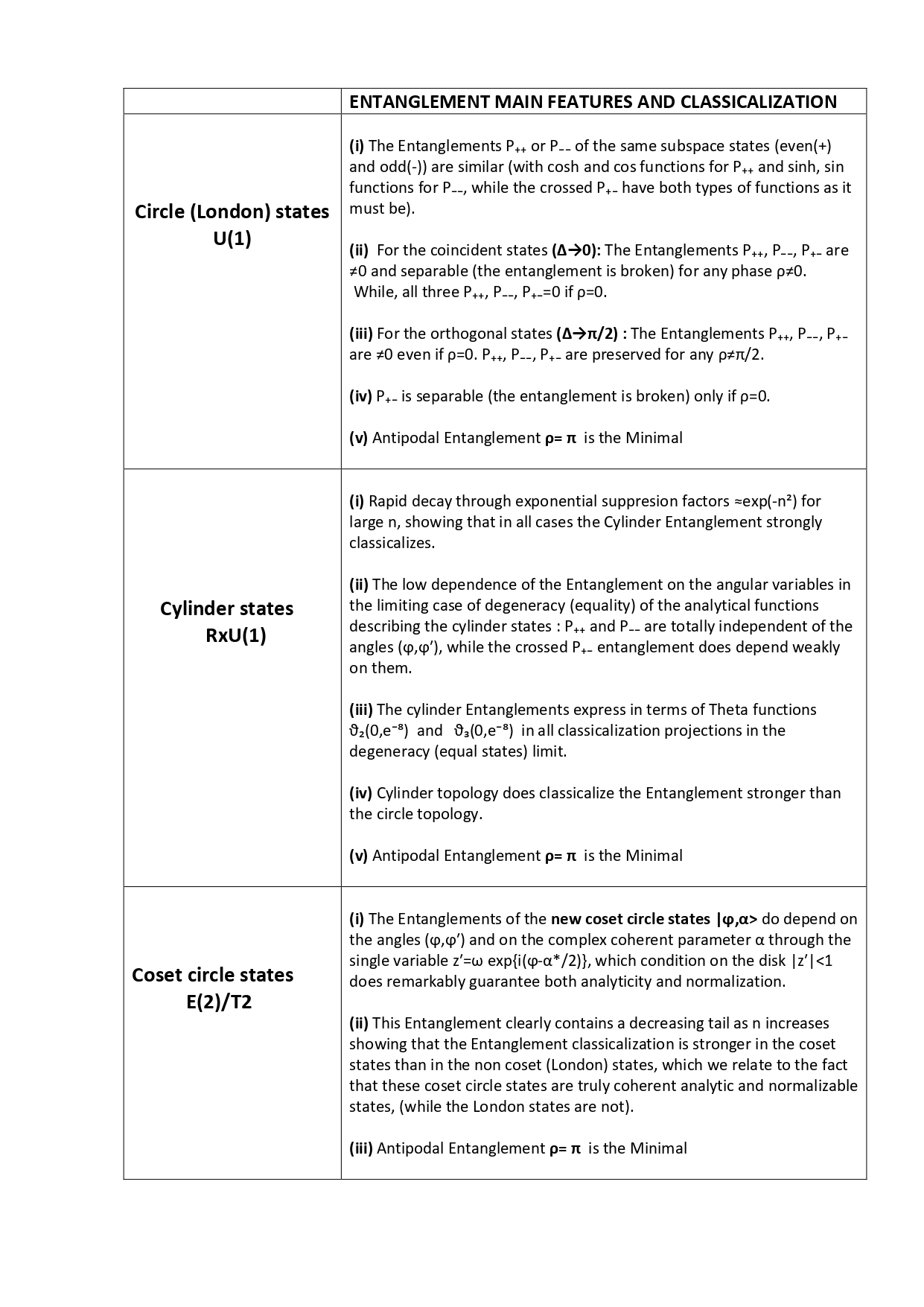}
\caption{
{\bf Some  Main New Features of Entanglements and Classicalization of this Paper. Sections I and VI provide more Summary and explanation.}}
\end{figure} 
\end{center}

\section{Entanglement of Coherent States in the Circle }

The classical states for the circle are coherent states in phase space, the
angle variable,(or so called London states). Let us consider two of these
states $\left\vert \varphi \right\rangle $, $\left\vert \varphi ^{\prime
}\right\rangle $ which in terms of the harmonic oscillator eigenstates $n,m$
are given by: 

\begin{equation}
\left\vert \,\varphi \,\right\rangle \;=\;\frac{1}{2\pi }\;\;\underset{%
n\,=\,0,\,1,\,2...}{{\displaystyle\sum }\; e^{-i\,\varphi \,n}}\,\left\vert
\,n\,\right\rangle  \label{a1}
\end{equation}%

\begin{equation}
\left\vert \,\varphi ^{\prime }\,\right\rangle \;=\;\frac{1}{2\pi }\;\;%
\underset{m\,=\,0,\,1,\,2...}{{\displaystyle\sum }\;e^{-i\,\varphi ^{\prime
}\,m}}\;\left\vert \,m\,\right\rangle  \label{a2}
\end{equation}

\bigskip

An entangled state can be defined naturally as%

\begin{equation}
\left\vert \,\Phi \,\right\rangle \;\equiv \;\frac{1}{\sqrt{2}}\;\left(
\,\left\vert \varphi \right\rangle \otimes \left\vert \varphi ^{\prime
}\right\rangle \,+\,e^{i\rho }\;\left\vert \varphi ^{\prime }\right\rangle
\otimes \left\vert \varphi \right\rangle \,\right)  \label{a3}
\end{equation}%
\\
where the order of the states in Eq. (\ref{a3}) indicates the
exchange of the angular coordinate (in this case $\varphi ,\varphi ^{\prime
} $ ), and the phase $\rho $ is an admissible control parameter which is very useful for the experimental realization of entanglement.

\medskip

Now, following the Minimal Group Representation consisting of the metaplectic group $Mp(2)$, we make the metaplectic projection of the entangled
state. To this end, we perform the $Mp(2)$ product state : 
$$\left\vert \,\Psi^{(\pm
)}\left(  \omega\right)  \Psi^{(\pm)}\left(  \sigma\right) \, \right \rangle \,\in \,
Mp\left(  2\right) ,  $$
\[
\left\vert \,\Psi^{(\pm)}\left(  \omega\right)  \Psi^{(\pm)}\left(
\sigma\right) \, \right\rangle \,= \,\frac{1}{\sqrt{2}} \left[\;  \left\vert \Psi
^{(\pm)}\left(  \omega\right)  \right\rangle \otimes\left\vert \Psi^{(\pm
)}\left(  \sigma\right)  \right\rangle +\left\vert \Psi^{(\pm)}\left(
\sigma\right)  \right\rangle \otimes\left\vert \Psi^{(\pm)}\left(
\omega\right)  \right\rangle \;\right]
\]
\\
The even $(+)$ and odd $(-)$ $Mp(2)$ states (irreducible Metaplectic representations) are given by: 
\begin{subequations}
\label{0}%
\begin{equation}%
\begin{array}
[c]{c}%
\\
\left\langle \,\varphi\right\vert \left\vert \Psi^{(\pm)}\left(
\omega\right)  \,\right\rangle =\\
\end{array}
\left\{
\begin{array}
[c]{cc}%
\left(  1-\left\vert \omega\right\vert ^{2}\right)  ^{1/4}\underset
{n\,=\,0,1,2..}{\sum}\frac{\left(  \omega e^{i\varphi}/2\right)  ^{2n}}{\sqrt
{2n!}} & \text{\ \ (+): \ even states}\\
& \\
\left(  1-\left\vert \omega\right\vert ^{2}\right)  ^{3/4}\underset
{n\,=\,0,1,2..}{\sum}\frac{\left(  \omega e^{i\varphi}/2\right)  ^{2n+1}}%
{\sqrt{\left(  2n+1\right)  !}} & \text{\ \ (-):\ \ odd states}%
\end{array}
\right.
\end{equation}
\\
Therefore, the total or complete projected state $\left\langle \;\varphi
\right\vert \left\vert \;\Psi\left(  \omega\right)  \;\right\rangle $ is given
by:
\end{subequations}
\begin{equation}
\left\langle \;\varphi\right\vert \left\vert \;\Psi\left(  \omega\right)
\;\right\rangle \;=\;\left\langle \;\varphi\right\vert \left\vert \;\Psi
^{(+)}\left(  \omega\right)  \;\right\rangle \;+\;\left\langle \;\varphi
\right\vert \left\vert \;\Psi^{(-)}\left(  \omega\right)  \;\right\rangle \;
\end{equation}

\begin{equation}
\left\langle \;\varphi\;\right\vert \left\vert \;\Psi\left(  \omega\right)
\;\right\rangle \;=\;\frac{\left(  1-\left\vert z\right\vert ^{2}\right)
^{1/4}}{\sqrt{2\pi}}\underset{n\,=\,0,1,2..}{\sum}\frac{\left(  z/2\right)
^{2n}}{\sqrt{(2n)\,!}}\left[  \,1\;+\;\left(  1-\left\vert z\right\vert
^{2}\right)  ^{1/2}\frac{\left(  z/2\right)  }{\sqrt{2n+1}}\;\right]
\label{f}%
\end{equation}
\\
where $z \;=\; \omega\,e^{i\varphi}.$ 

Now, having the above expressions into account, we define in the Bargmann representation
the following functions in order to perform the respective projections:%
\begin{align*}
z_{1}  &  \equiv\omega e^{i\varphi}\rightarrow\left\langle \,\varphi
\right\vert \left\vert \Psi^{(\pm)}\left(  \omega\right)  \,\right\rangle
\equiv\Psi^{(\pm)}\left(  z_{1}\right),\text{  }
&  \text{\ \ \  }
\text{\ }z_{1}^{\prime}  &  \equiv\omega e^{i\varphi^{\prime}}\rightarrow
\left\langle \,\varphi^{\prime}\right\vert \left\vert \Psi^{(\pm)}\left(
\omega\right)  \,\right\rangle \equiv\Psi^{(\pm)}\left(  z_{1}^{\prime}\right)
\\
& \\
z_{2}  &  \equiv\sigma e^{i\varphi}\,\rightarrow\left\langle \,\varphi
\right\vert \left\vert \Psi^{(\pm)}\left(  \sigma\right)  \,\right\rangle
\equiv\Psi^{(\pm)}\left(  z_{2}\right), 
& 
\text{\ \ }z_{2}^{\prime}  &  \equiv\sigma e^{i\varphi^{\prime}}%
\;\rightarrow\left\langle \,\varphi^{\prime}\right\vert \left\vert \Psi^{(\pm
)}\left(  \sigma\right)  \,\right\rangle \equiv\Psi^{(\pm)}\left(
z_{2}^{\prime}\right)
\end{align*}
Therefore :%
\[
\left\langle \Psi^{(\pm)}\left(  \omega\right)  \Psi^{(\pm)}\left(
\sigma\right)  \right\vert \left\vert \Phi\right\rangle \;=\;\frac{1}{2}\left[
\Psi^{(\pm)}\left(  z_{1}\right)  \otimes\Psi^{(\pm)}\left(  z_{2}^{\prime
}\right)  \;+\;e^{i\rho}\Psi^{(\pm)}\left(  z_{1}^{\prime}\right)  \otimes
\Psi^{(\pm)}\left(  z_{2}\right)  \right]
\]

Or, due to Eq. (\ref{f}) using the complete projected states, we have:%

\begin{align}
\left\langle \Psi\left(  \omega\right)  \Psi\left(  \sigma\right)  \right\vert
\left\vert \Phi\right\rangle  &  =\frac{1}{2}\left(  1-\left\vert
\omega\right\vert ^{2}\right)  ^{1/4}\left(  1-\left\vert \sigma\right\vert
^{2}\right)  ^{1/4}\underset{n,m=0,1,2...}{%
{\displaystyle\sum}
}\frac{\left(  \omega^{\ast}/2\right)  ^{2n}}{\sqrt{2n!}}\frac{\left(
\sigma^{\ast}/2\right)  ^{2m}}{\sqrt{2m!}}\times\label{gwf}\\
&  \times\left[  e^{-2i\left(  \varphi n+\varphi^{\prime}m\right)  }\left(
1+\left(  1-\left\vert \omega\right\vert ^{2}\right)  ^{1/2}\frac{\left(
\omega^{\ast}e^{-i\varphi}/2\right)  }{\sqrt{2n+1}}\right)  \left(  1+\left(
1-\left\vert \sigma\right\vert ^{2}\right)  ^{1/2}\frac{\left(  \sigma^{\ast
}e^{-i\varphi^{\prime}}/2\right)  }{\sqrt{2m+1}}\right)  \right.  +\nonumber\\
&  \left.  e^{i\rho}e^{-2i\left(  \varphi^{\prime}n+\varphi m\right)  }\left(
1+\left(  1-\left\vert \omega\right\vert ^{2}\right)  ^{1/2}\frac{\left(
\omega^{\ast}e^{-i\varphi^{\prime}}/2\right)  }{\sqrt{2n+1}}\right)  \left(
1+\left(  1-\left\vert \sigma\right\vert ^{2}\right)  ^{1/2}\frac{\left(
\sigma^{\ast}e^{-i\varphi}/2\right)  }{\sqrt{2m+1}}\right)  \right] \nonumber
\end{align}

\subsection{Entanglement and Minimal Group Representation}

The minimal group representation which is provided by the metaplectic group implies for  the circle states the crossed even (+)  odd (-)  explicit projection :

\begin{gather*}
\left\langle \Psi^{(+)}\left(  \omega\right)  \Psi^{(-)}\left(  \omega\right)
\right\vert \left\vert \Phi\right\rangle =\frac{1}{2}\left(  1-\left\vert
\omega\right\vert ^{2}\right) \\
\underset{n,\,m\,=\,0,1,2...}{%
{\displaystyle\sum}
}\left[  \frac{\left(  \omega^{\ast}/2\right)  ^{2(n+m)+1}}{\sqrt{\left(
2n\right)  !\left(  2m+1\right)  !}}\left(  e^{-i2n\varphi}e^{-i\left(
2m+1\right)  \varphi^{\prime}}+e^{i\rho}e^{-i2n\varphi^{\prime}}e^{-i\left(
2m+1\right)  \varphi}\right)  \right]
\end{gather*}%
\\
and its square norm Probability P$_{+-}$\;: 
\begin{gather*}
P_{+-} = \left\Vert \left\langle \Psi^{(+)}\left(  \omega\right)  \Psi^{(-)}\left(
\omega\right)  \right\vert \left\vert \Phi\right\rangle \right\Vert ^{2}%
\;= \;\frac{1}{2}\left(  1-\left\vert \omega\right\vert ^{2}\right)  ^{2}\left\{
\cosh\left(  \frac{\left\vert \omega\right\vert ^{2}}{4}\right)  \sinh\left(
\frac{\left\vert \omega\right\vert ^{2}}{4}\right)  + \right. \\
\left.  +\cosh\left(  \frac{\left\vert \omega\right\vert ^{2}}{4}\cos
\Delta\right)  \sinh\left(  \frac{\left\vert \omega\right\vert ^{2}}{4}%
\cos\Delta\right)  \cos\rho + \cos\left(  \frac{\left\vert \omega\right\vert
^{2}}{4}\sin\Delta\right)  \sin\left(  \frac{\left\vert \sigma\right\vert
^{2}}{4}\sin\Delta\right)  \sin\rho\right\}
\end{gather*}

\bigskip

In what follows, we analyze the particular projections in each sector
(even and odd) of the entangled wave functions together with the
corresponding square norm Probabilities 

\bigskip

\subsection{Entanglement Probabilities of the Even (+) and Odd (-) Hilbert space sectors}

Considering that the basic states of $Mp(n)$ solve the identity in each irreducible subspace representation 
$s\, = \, 1/4,\, 3/4$, of the total Hilbert space, we will show the corresponding
expressions of the {\bf Entanglement Probabilities} from the general wave function
Eq. (\ref{gwf}) projected in each subspace for simplicity

\bigskip

{\bf - Entanglement Probability $P_{++}$ of the Even (++) Sectors: }

\bigskip

The explicit projected (+)(+) entangled expression of the London states wave function is:
\begin{gather*}
\left\langle \Psi^{(+)}\left(  \omega\right)  \Psi^{(+)}\left(  \sigma\right)
\right\vert \left\vert \Phi\right\rangle \;= \; 
\frac{1}{2}\left(  1-\left\vert
\omega\right\vert ^{2}\right)  ^{1/4}\left(  1-\left\vert \sigma\right\vert
^{2}\right)  ^{1/4}\\
\underset{n,\,m\;=\;\,0,\,1,\,2...}{%
{\displaystyle\sum}
}\left[\;  \frac{\left(  \omega^{\ast}e^{-i\varphi}/2\right)  ^{2n}}{\sqrt{2n!}%
}\frac{\left(  \sigma^{\ast}e^{-i\varphi^{\prime}}/2\right)  ^{2m}}{\sqrt
{2m!}}\;+\;\frac{\left(  \omega^{\ast}e^{-i\varphi^{\prime}}/2\right)  ^{2n}%
}{\sqrt{2n!}}\frac{\left(  \sigma^{\ast}e^{-i\varphi}/2\right)  ^{2m}}%
{\sqrt{2m!}}\;e^{i\rho}\;\right]
\label{++}
\end{gather*}

\bigskip

Defining $\Delta \equiv \left( \varphi -\varphi ^{\prime }\right)$  in the
limit for the same variable : $\varphi \rightarrow \varphi ^{\prime
},\; ie\;\,\Delta \rightarrow 0$,\, and for the orthogonal states, ie\ $\Delta
\rightarrow \;\pi /2$, the entanglement Probabilities $P_{++}$ (strictly the norm
square) considerably simplify
as given explicitly in the Appendix I whose limits of interest $\Delta
\rightarrow 0$ (complete degeneration) and $\Delta \rightarrow \pi /2$ are
the following:

\bigskip

\[
\underset{\,\Delta\rightarrow \;0}{\lim}\,P_{++}\;=\;\frac{1}{2}\sqrt{\left(  1-\left\vert
\omega\right\vert ^{2}\right)  \left(  1-\left\vert \sigma\right\vert
^{2}\right)  }\cosh\left(  \frac{\left\vert \omega\right\vert ^{2}}{4}\right)
\cosh\left(  \frac{\left\vert \sigma\right\vert ^{2}}{4}\right)  \left(
1-\cos\rho\right)
\]%
\[
\underset{\,\Delta\rightarrow\pi/2}{\lim}\,P_{++}\; =\; \frac{1}{2}\sqrt{\left(  1-\left\vert
\omega\right\vert ^{2}\right)  \left(1-\left\vert \sigma\right\vert
^{2}\right)  }\left\{  \cosh\left(  \frac{\left\vert \omega\right\vert ^{2}%
}{4}\right)  \cosh\left(  \frac{\left\vert \sigma\right\vert ^{2}}{4}\right)
+\cos\left(  \frac{\left\vert \omega\right\vert ^{2}}{4}\right)  \cos\left(
\frac{\left\vert \sigma\right\vert ^{2}}{4}\right)  \cos\rho\right\}
\]
\\
We see the role played by the phase $\rho$ in the entanglement $P_{++}$ which makes it non zero iff $\rho \neq 0$ for the case of the coincident states $\,(\Delta\rightarrow \,0)$, while the entanglement is non-zero even if  $\rho = 0 $ for the orthogonal states $(\,\Delta\rightarrow\,\pi/2)$.

\medskip

For $\rho = 0$,\; we see that\,: 
$$ 
\underset{\,\Delta \rightarrow \; 0}{\lim}\;P_{++}\;=\;0,$$\;\;
{and} \;$$\underset{\,\Delta \rightarrow \;\pi /2}{\lim}\,P_{++}\;=\;\frac{1}{2}\sqrt{%
\left( 1-\left\vert \omega \right\vert ^{2}\right) \left( 1-\left\vert
\sigma \right\vert ^{2}\right) }\cosh \left( \frac{\left\vert \omega
\right\vert ^{2}}{4}\right) \cosh \left( \frac{\left\vert \sigma \right\vert
^{2}}{4}\right) 
$$\;
Therefore, the {\it state is separable}, that is to say, the {\it entanglement is broken} in this case.

\bigskip

{\bf - Entanglement Probability $P_{+\,-}$ of the Even-Odd (+\,-) Sectors:}

\bigskip

The London states  crossed projected into the even (+) and odd (-) Metaplectic states have the following explicit expression:
\begin{gather*}
\left\langle \Psi^{(+)}\left(  \omega\right)  \Psi^{(-)}\left(  \sigma\right)
\right\vert \left\vert \Phi\right\rangle\, = \,\frac{1}{2}\left(  1-\left\vert
\omega\right\vert ^{2}\right)^{1/4}\left(  1-\left\vert \sigma\right\vert
^{2}\right)  ^{3/4}\\
\underset{n,\;m\; =\; 0\,,1\,,2\,...}{%
{\displaystyle\sum}
}\left[\;  \frac{\left(  \omega^{\ast}e^{-i\varphi}/2\right)  ^{2n}}%
{\sqrt{\left(  2n\right)  !}}\;\frac{\left(  \sigma^{\ast}e^{-i\varphi^{\prime}%
}/2\right)^{2m+1}}{\sqrt{\left( 2m+1\right)  !}}\,+\,e^{i\rho}\;\frac{\left(
\omega^{\ast}e^{-i\varphi^{\prime}}/2\right)  ^{2n}}{\sqrt{2n!}}\;\frac{\left(
\sigma^{\ast}e^{-i\varphi}/2\right) ^{2m+1}}{\sqrt{\left(2m+1\right)  !}%
}\;\right]
\end{gather*}%

\bigskip

\medskip

Now we see the role played by the phase $\rho $ in the entanglement probability $%
P_{+\,-}$ in the corresponding limits:%

\bigskip

{\bf In the coincident limit} :
\begin{equation*}
\underset{\Delta \;\rightarrow \; 0}{\lim}\;P_{+\,-}\;= \;\frac{1}{2}\left(
1-\left\vert \omega \right\vert ^{2}\right) ^{1/2}\left( 1-\left\vert \sigma
\right\vert ^{2}\right) ^{3/2}\cosh \left( \frac{\left\vert \omega
\right\vert ^{2}}{4}\right) \sinh \left( \frac{\left\vert \sigma \right\vert
^{2}}{4}\right) \left( 1-\cos \rho \right)
\end{equation*}%

We see the above expression is separable (the entanglement is broken) for any value of the phase $\rho\; \ne \; 0$ while $\underset{\Delta \rightarrow \,0}{\lim}\,P_{+\,-} = \,0$ \
for $\rho = 0.$

\medskip

\bigskip

{\bf In the orthogonal limit} : 
\begin{equation*}
\underset{\Delta \rightarrow \,\pi /2}{\lim}\;P_{+-} = \frac{1}{2}\left(
1-\left\vert \omega \right\vert ^{2}\right) ^{1/2}\left( 1-\left\vert \sigma
\right\vert ^{2}\right) ^{3/2}\left\{ \cosh \left( \frac{\left\vert \omega
\right\vert ^{2}}{4}\right) \sinh \left( \frac{\left\vert \sigma \right\vert
^{2}}{4}\right) +\cos \left( \frac{\left\vert \omega \right\vert ^{2}}{4}%
\right) \sin \left( \frac{\left\vert \sigma \right\vert ^{2}}{4}\right) \sin
\rho \right\}
\end{equation*}

In this case the entanglement is preserved for any value of $\rho \neq 0.$ Only it is separable (the entanglement is broken) for $\rho = 0$.

\bigskip

\bigskip

{\bf - Entanglement Probability $P_{--}$ of the Odd (-\,-) Sectors } : %

\bigskip

For the odd Hilbert space sector, the projection of the circle states is given by:
\begin{gather*}
\left\langle \Psi^{(-)}\left(  \omega\right)  \Psi^{(-)}\left(  \sigma\right)
\right\vert \left\vert \Phi\right\rangle \;= \;\frac{1}{2}\left(  1-\left\vert
\omega\right\vert ^{2}\right)  ^{3/4}\left(  1-\left\vert \sigma\right\vert
^{2}\right)  ^{3/4}\\
\underset{n,\;m\;=\;0,\;1,\;2\;...}{%
{\displaystyle\sum}
}\left[\;\frac{\left(  \omega^{\ast}e^{i\varphi}/2\right)  ^{2n+1}}%
{\sqrt{\left(  2n+1\right)  !}}\;\frac{\left(  \sigma^{\ast}e^{i\varphi^{\prime
}}/2\right)  ^{2m+1}}{\sqrt{\left(  2m+1\right)  !}}\;+\;\frac{\left(
\omega^{\ast}e^{i\varphi^{\prime}}/2\right)  ^{2n+1}}{\sqrt{\left(
2n+1\right)  !}}\;\frac{\left(  \sigma^{\ast}e^{i\varphi}/2\right)  ^{2m+1}%
}{\sqrt{\left(  2m+1\right)  !}}e^{i\rho}\;\right]
\end{gather*}

\bigskip

Now, we see the role played by the phase $\rho$ in the entanglement probability  $P_{--}$ from the corresponding limits: 
 
 \bigskip

{\bf In the coincident limit}:
\begin{equation*}
\underset{\Delta \;\rightarrow \; 0}{\lim}\;P_{-\,-}\;= \;\frac{1}{2}\left(
1-\left\vert \omega \right\vert ^{2}\right) ^{1/2}\left( 1-\left\vert \sigma
\right\vert ^{2}\right) ^{3/2}\sinh \left( \frac{\left\vert \omega
\right\vert ^{2}}{4}\right) \sinh \left( \frac{\left\vert \sigma \right\vert
^{2}}{4}\right) \left( 1-\cos \rho \right)
\end{equation*}%

In the limit of coincident states $P_{--}$ is separable  (the entanglement is broken) for any value of the phase  $\rho \ne 0 $,  while it is vanishes for $\rho = 0$ : $$
\underset{\Delta \;\rightarrow \;0}{\lim}\;P_{--}\; = \; 0 \;\;\;
\text{when}\;\; \rho \;=\; 0
$$

{\bf In the orthogonal limit}:
\begin{equation*}
\underset{\Delta\; \rightarrow \;\pi /2}{\lim}P_{--}=\frac{1}{2}\left[ \left(
1-\left\vert \omega \right\vert ^{2}\right) \left( 1-\left\vert \sigma
\right\vert ^{2}\right) \right] ^{3/2}\left\{ \sinh \left( \frac{\left\vert
\omega \right\vert ^{2}}{4}\right) \sinh \left( \frac{\left\vert \sigma
\right\vert ^{2}}{4}\right) +\sin \left( \frac{\left\vert \omega \right\vert
^{2}}{4}\right) \sin \left( \frac{\left\vert \sigma \right\vert ^{2}}{4}%
\right) \cos \rho \right\}
\end{equation*}%

In this case the entanglement is preserved for any value of $\rho \neq \pi /2$ (within
the respective modulus), while  the entanglement is broken for $\rho \;=\;\pi /2$.

\medskip

It must be noticed, and as it can be seen in Figure 1: For the coincident states $\Delta = 0$, the {\bf antipodal entanglement} (regulated by the control parameter $\rho =\pi $) generates  a effect of {\bf minimum
 entanglement} probability in {\bf all} the Mp(2) projections . This is in marked
contrast to the entanglement of orthogonal states  $\Delta =\pi /2$ in Figure \ref{f2} where the antipodal
entanglement ($\rho =\pi $) is relevant, inverting the maxima of the probability.

\subsection{ Full Entangled Probability}

For completeness, to calculate the square norm in the full entangled state, we
introduce a polar decomposition for the functions of the states of Mp(2),
namely%
\[
\omega=\left\vert \omega\right\vert e^{i\theta_{1}},\text{
\ \ \ \ \ \ \ \ \ \ \ \ \ \ \ \ }\sigma=\left\vert \sigma\right\vert
e^{i\theta_{2}}%
\]
Then:%
\begin{gather*}
\left\Vert \left\langle \Psi\left(  \omega\right)  \Psi\left(  \sigma\right)
\right\vert \left\vert \Phi\right\rangle \right\Vert ^{2}=\frac{1}{4}\left(
1-\left\vert \omega\right\vert ^{2}\right)  ^{1/2}\left(  1-\left\vert
\sigma\right\vert ^{2}\right)  ^{1/2}\underset{n,\,m\,=\,0,1,2...}{%
{\displaystyle\sum}
}\frac{\left(  \left\vert \omega\right\vert ^{2}/4\right)  ^{2n}}{2n!}%
\frac{\left(  \left\vert \sigma\right\vert ^{2}/4\right)  ^{2m}}{2m!}\times\\
\times\left(  1+\left(  1-\left\vert \omega\right\vert ^{2}\right)
^{1/2}\frac{\cos\left(  \theta_{1}+\varphi\right)  }{\sqrt{2n+1}}\left\vert
\omega\right\vert +\frac{\left(  1-\left\vert \omega\right\vert ^{2}\right)
}{4\left(  2n+1\right)  }\left\vert \omega\right\vert ^{2}\right) \\
\left(  1+\left(  1-\left\vert \sigma\right\vert ^{2}\right)  ^{1/2}\frac
{\cos\left(  \theta_{2}+\varphi^{\prime}\right)  }{\sqrt{2m+1}}\left\vert
\sigma\right\vert +\frac{\left(  1-\left\vert \sigma\right\vert ^{2}\right)
}{4\left(  2m+1\right)  }\left\vert \sigma\right\vert ^{2}\right)  +\\
+\left(  1+\left(  1-\left\vert \omega\right\vert ^{2}\right)  ^{1/2}%
\frac{\cos\left(  \theta_{1}+\varphi^{\prime}\right)  }{\sqrt{2n+1}}\left\vert
\omega\right\vert +\frac{\left(  1-\left\vert \omega\right\vert ^{2}\right)
}{4\left(  2n+1\right)  }\left\vert \omega\right\vert ^{2}\right) \\
\left(  1+\left(  1-\left\vert \sigma\right\vert ^{2}\right)  ^{1/2}\frac
{\cos\left(  \theta_{2}+\varphi\right)  }{\sqrt{2m+1}}\left\vert
\sigma\right\vert +\frac{\left(  1-\left\vert \sigma\right\vert ^{2}\right)
}{4\left(  2m+1\right)  }\left\vert \sigma\right\vert ^{2}\right)  +\\
+\cos\left(  \rho+\left(  \varphi^{\prime}-\varphi\right)  (n-m\right)
)\;(\,AB-CD\,)
\end{gather*}
where\;:%
\begin{align*}
A  &  \;=\;\frac{1}{2}\left[\;  2+\left(  1-\left\vert \omega\right\vert
^{2}\right)  ^{1/2}\frac{cos\left(  \theta_{1}+\varphi\right)  }{\sqrt{2n+1}%
}\left\vert \omega\right\vert +\left(  1-\left\vert \sigma\right\vert
^{2}\right)  ^{1/2}\frac{\cos\left(  \theta_{2}+\varphi\right)  }{\sqrt{2m+1}%
}\left\vert \sigma\right\vert +\right. \\
&  \left.  \frac{\left(  1-\left\vert \sigma\right\vert ^{2}\right)
^{1/2}\left(  1-\left\vert \omega\right\vert ^{2}\right)  ^{1/2}}{\sqrt
{2n+1}\;\sqrt{2m+1}}\frac{\left\vert \omega\right\vert \left\vert \sigma
\right\vert }{2}\cos\left(  \theta_{1}-\theta_{2}\right) \; \right] \\
& \\
B  &  \;=\;\frac{1}{2}\left[ \; 2-\left(  1-\left\vert \omega\right\vert
^{2}\right)  ^{1/2}\frac{\cos\left(  \theta_{1}+\varphi^{\prime}\right)
}{\sqrt{2n+1}}\left\vert \omega\right\vert +\left(  1-\left\vert
\sigma\right\vert ^{2}\right)  ^{1/2}\frac{\cos\left(  \theta_{2}%
+\varphi^{\prime}\right)  }{\sqrt{2m+1}}\left\vert \sigma\right\vert +\right.
\\
&  \left.  \frac{\left(  1-\left\vert \sigma\right\vert ^{2}\right)
^{1/2}\left(  1-\left\vert \omega\right\vert ^{2}\right)  ^{1/2}}{\sqrt
{2n+1}\sqrt{2m+1}}\frac{\left\vert \omega\right\vert \left\vert \sigma
\right\vert }{2}\cos\left(  \theta_{2}-\theta_{1}\right)  \right]
\end{align*}%
\begin{align*}
C  & \; =\;\frac{1}{2}\left[\;  \left(  1-\left\vert \omega\right\vert ^{2}\right)
^{1/2}\frac{\sin\left(  \theta_{1}+\varphi\right)  }{\sqrt{2n+1}}\left\vert
\omega\right\vert -\left(  1-\left\vert \sigma\right\vert ^{2}\right)
^{1/2}\frac{\sin\left(  \theta_{2}+\varphi\right)  }{\sqrt{2m+1}}\left\vert
\sigma\right\vert +\right. \\
&  \left.  \frac{\left(  1-\left\vert \sigma\right\vert ^{2}\right)
^{1/2}\left(  1-\left\vert \omega\right\vert ^{2}\right)  ^{1/2}}{\sqrt
{2n+1}\sqrt{2m+1}}\frac{\left\vert \omega\right\vert \left\vert \sigma
\right\vert }{2}\sin\left(  \theta_{1}-\theta_{2}\right) \; \right] \\
& \\
D  & \; = \;\frac{1}{2}\left[ \; -\left(  1-\left\vert \omega\right\vert ^{2}\right)
^{1/2}\frac{\sin\left(  \theta_{1}+\varphi^{\prime}\right)  }{\sqrt{2n+1}%
}\left\vert \omega\right\vert +\left(  1-\left\vert \sigma\right\vert
^{2}\right)  ^{1/2}\frac{\sin\left(  \theta_{2}+\varphi^{\prime}\right)
}{\sqrt{2m+1}}\left\vert \sigma\right\vert +\right. \\
&  \left.  \frac{\left(  1-\left\vert \sigma\right\vert ^{2}\right)
^{1/2}\left(  1-\left\vert \omega\right\vert ^{2}\right)  ^{1/2}}{\sqrt
{2n+1}\sqrt{2m+1}}\frac{\left\vert \omega\right\vert \left\vert \sigma
\right\vert }{2}\sin\left(  \theta_{2}-\theta_{1}\right) \; \right]
\end{align*}

\bigskip

As we can see, in the general case the total output of the London
entanglement states is evidently manifested with the highest degree of entanglement.

\begin{figure}
[ptb]
\begin{center}
\includegraphics[
height=6.499000in,
width=6.120200in,
height=6.4022in,
width=6.3648in
]
{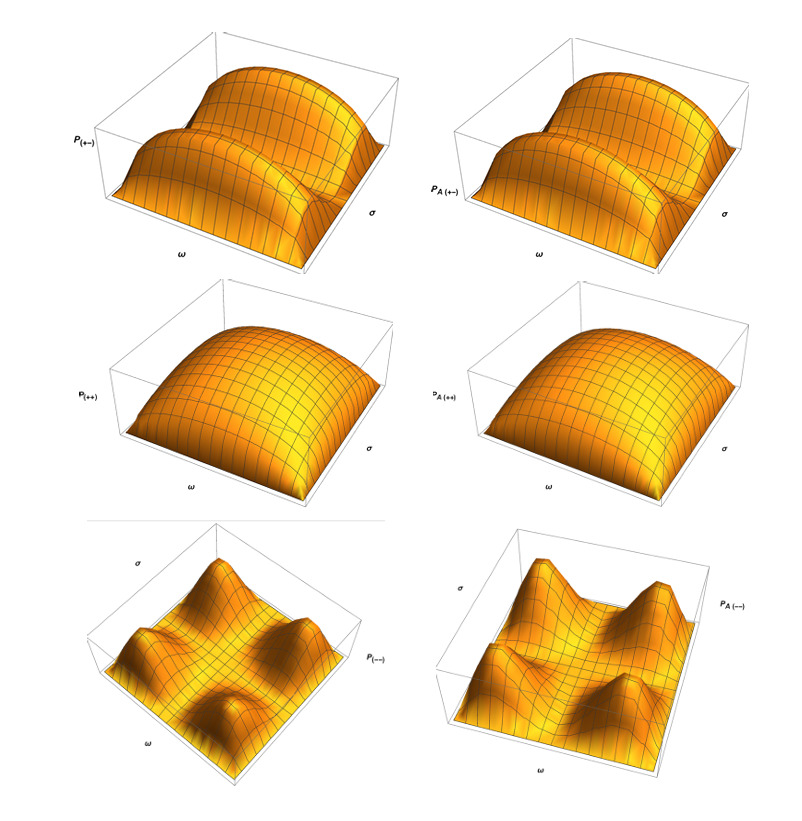}
\caption{{\bf Probabilities of
the Entanglement} of two circle (London) states with  $\Delta = 0$ ($\protect\varphi = %
\protect\varphi^{\prime }$): {\bf coincident states} projected onto the Metaplectic group (Minimal Representation Group): {\bf Classicalization}: Left side with the control entanglement phase  $\protect\rho = 0$. Right side with $%
\protect\rho =\protect\pi $: {e.g \bf Antipodal Entanglement} . In this case, {\bf the Antipodal}
condition (regulated by the control parameter $\protect\rho = \protect\pi $)
does not modify essentially  any of the Entanglement Mp (2) Classicalizations.}%
\label{f1}%
\end{center}
\end{figure} 
\begin{figure}
[ptb]
\begin{center}
\includegraphics[
height=6.499000in,
width=6.120200in,
]
{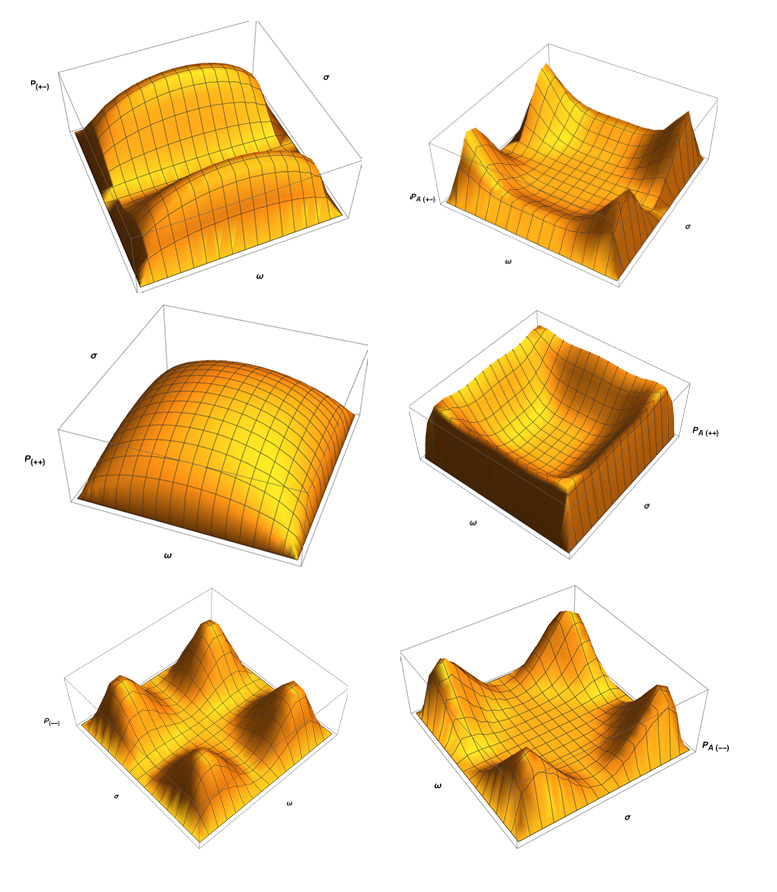}
\caption{{\bf Probabilities of the Entanglement} of two circle London states $%
\Delta =\protect\pi /2$ ($\protect\varphi =\protect\varphi ^{\prime }$ +$%
\protect\pi /2$): {\bf orthogonal states}  projected onto the Metaplectic group (Minimal Representation Group): {\bf Classicalization}. Left side entanglement with the control paramenter phase $\protect\rho = 0$. Right side with $\protect\rho = %
\protect\pi $: {\bf Antipodal Entanglament}. Here the antipodal condition is relevant,
inverting the maxima of the Entanglement Probabilities.}%
\label{f2}%
\end{center}
\end{figure} 

\section{Entanglement of Coherent States in the Cylinder}

As we pointed out recently in  Ref \cite{APL25} in order to introduce the Coherent States for a
quantum particle on the cylinder geometry, it is possible to follow the
Barut--Girardello construction and seek the Coherent State as the solution of
the eigenvalue equation:
\[
X \left\vert \,\xi \,\right\rangle \;=\;\xi|\left\vert \,\xi\,\right\rangle
\]

with the complex parameter $\xi$, similarly to the standard case, and
\[
X\;=\;e^{\,i\,\left(  \widehat{\varphi}\,+\,\widehat{J}\right)  }%
\]
In order to analyze the Coherent State of a particle in the cylinder in the
context of the Minimal Group Representation, we express the coherent states as:%

\begin{equation}
\label{cilins}\left\vert \,\xi\,\right\rangle \;=\underset{}{\underset
{j\;=\;-\infty}{\overset{\infty}{\sum}}e^{\left( \, l-i\,\varphi\,\right)
j}\,e^{-j^{2}/2}}\,\left\vert \, j \,\right\rangle
\end{equation}

{\bf (i)} If \;$\left\vert \,j\,\right\rangle $ $\sim$ $\left\vert \,n\,\right\rangle
$\; , and for the Metaplectic 
$s\;=\;1/4$ states $\left\vert \,\Psi^{\left(
+\right)  }\left(  \omega\right)  \, \right\rangle $ : \;%

\begin{align*}
\left\langle \, \xi\,\right\vert \left\vert \,\Psi^{\left(  +\right)  }\left(
\omega\right)  \, \right\rangle  &  =\left(  1-\left\vert \omega\right\vert
^{2}\right)  ^{1/4}\overset{\infty}{\underset{m\;=\;-\infty\;}{\,\sum}%
}\underset{\,n\;=\;0,1,2...}{\;\sum}\frac{\left(  \omega/2\right)  ^{2n}%
}{\sqrt{2n\,!}}\;e^{\left(  l-i\varphi\right)  \, m}\,e^{-2m^{2}}\left\langle
m\right\vert \left\vert 2n\right\rangle \\
\\
&  = \,\left(  1-\left\vert \omega\right\vert ^{2}\right)  ^{1/4}%
\underset{n\;=\;0,1,2..}{\sum}\frac{\left(  \, \omega\, e^{\,\left(
l-i\varphi\right)  }/2\right)  ^{2n}}{\sqrt{2n\,!}}\,e^{-2n^{2}}%
\end{align*}

\bigskip

{\bf (ii)} Similarly, for the $s\;=\;3/4$ Metaplectic states $\left\vert \,\Psi^{\left(
- \right)  }\left(  \omega\right)  \, \right\rangle $ :
\begin{align*}
\left\langle \,\xi\,\right\vert \left\vert \,\Psi^{\left(  -\right)  }\left(
\omega\right)  \, \right\rangle  &  = \left(  1-\left\vert \omega\right\vert
^{2}\right)  ^{3/4}\overset{\infty\;}{\underset{m\;=-\infty\;}{\sum}}%
\underset{n\;=\;0,1,2...}{\sum}\frac{\left(  \omega/2\right)  ^{2n+1}}%
{\sqrt{\left(  2n+1\right)  \, !}}\;e^{\left(  l-i\varphi\right)  \,m}\,
e^{-m^{2}/2}\,\left\langle m\right\vert \left\vert 2n+1\right\rangle \\
\\
&  = \left(  1 - \left\vert \omega\right\vert ^{2}\right)  ^{3/4}%
\underset{n\;=\;0,1,2...}{\sum}\frac{\left(  \, \omega\,e^{\left(
l-i\varphi\right)  }/2\,\right)  ^{2n+1}}{\sqrt{\left(  2n+1\right)  \,!}%
}e^{-\left(  2n+1\right)  ^{2}/2}%
\end{align*}

\bigskip

We see that the scalar product projections taken with the cilinder
$\left\langle \, \xi\,\right\vert $ space configuration states are similar to
the projections taken with the circle $\left\langle \, \varphi\,\right\vert $
phase space states, but \textit{in contrast} they contain weight functions:
$e^{-2n^{2}}$ and $e^{-\left(  2n+1\right)  ^{2}/2}$, which
\textit{drastically attenuate} the scalar products when $n\rightarrow\infty$:%

\begin{subequations}
\label{0}%
\begin{equation}%
\begin{array}
[c]{c}%
\\
\left\langle \,\xi\,\right\vert \left\vert \,\Psi^{(\,\pm)}\left(
\omega\right)  \, \right\rangle =
\end{array}
\left\{
\begin{array}
[c]{cc}%
\left(  1-\left\vert \omega\right\vert ^{2}\right)  ^{1/4}\underset
{n\;=\;0,1,2...}{\sum}\frac{\left(  \, \omega\,e^{\left(  l-i\varphi\right)
}/2\,\right)  ^{2n}}{\sqrt{2n\,!}}\,e^{\,-2n^{2}} & \text{even states}\\
& \\
\left(  1-\left\vert \omega\right\vert ^{2}\right)  ^{3/4}\underset
{n\;=\;0,1,2...}{\sum}\frac{\left(  \, \omega\,e^{\,\left(  l-i\varphi\right)
}/2\,\right)  ^{2n+1}}{\sqrt{\left(  2n+1\right)  \,!}}\;e^{-\,\left(
2n+1\right)  ^{2}/2} & \text{odd states}%
\end{array}
\right.
\end{equation}

\bigskip

Consequently, for the {\bf total state} :
\end{subequations}
\begin{equation}
\left\vert \,\Psi\left(  \omega\right)  \,\right\rangle \;=\;\left\vert
\,\Psi^{(+)}\left(  \omega\right)  \,\right\rangle \;+\;\left\vert
\,\Psi^{\,(-)}\left(  \omega\right)  \,\right\rangle , \label{total}%
\end{equation}
we have:%
\begin{equation}
\left\langle \,\xi\,\right\vert \left\vert \,\Psi\left(  \omega\right)
\,\right\rangle =\left(  1-\left\vert \omega\right\vert ^{2}\right)
^{1/4}\underset{n\;=\;0,1,2...}{\sum}\frac{\left(  \omega\,e^{\left(
l-i\varphi\right)  }/2\right)  ^{2n}}{\sqrt{2n\,!}}\,e^{-2\,n^{2}}\left[
1+\left(  1-\left\vert \omega\right\vert ^{2}\right)  ^{1/2}\,\frac
{\omega\,e^{\,\left(  l-i\varphi\right)  }}{\sqrt{2n+1}}\;e^{-(2n+1/2)}%
\,\right]  \label{wfcs}%
\end{equation}

In \ order to simplify, in analogy with the previous case, we define in the
Bargmann representation
\begin{align*}
z_{1}  &  \equiv\omega e^{\left(  l+i\varphi\right)  }\rightarrow\left\langle
\,\xi\right\vert \left\vert \Psi^{(\pm)}\left(  \omega\right)  \,\right\rangle
\equiv\Psi^{(\pm)}\left(  z_{1}\right),\text{ }
&  \text{}
\text{\ }z_{1}^{\prime}  &  \equiv\omega e^{\left(  l^{\prime}+i\varphi
^{\prime}\right)  }\rightarrow\left\langle \,\xi^{\prime}\right\vert
\left\vert \Psi^{(\pm)}\left(  \omega\right)  \,\right\rangle \equiv\Psi
^{(\pm)}\left(  z_{1}^{\prime}\right) \\
& \\
z_{2}  &  \equiv\sigma e^{\left(  l+i\varphi\right)  },\rightarrow\left\langle
\,\xi\right\vert \left\vert \Psi^{(\pm)}\left(  \sigma\right)  \,\right\rangle
\equiv\Psi^{(\pm)}\left(  z_{2}\right), 
& 
\text{\ \ }z_{2}^{\prime}  &  \equiv\sigma e^{\left(  l^{\prime}%
+i\varphi^{\prime}\right)  }\rightarrow\left\langle \,\xi^{\prime}\right\vert
\left\vert \Psi^{(\pm)}\left(  \sigma\right)  \,\right\rangle \equiv\Psi
^{(\pm)}\left(  z_{2}^{\prime}\right)
\end{align*}

Then, the wave entangled projected state is expressed as :%
\[
\left\langle \Psi^{(\pm)}\left(  \omega\right)  \Psi^{(\pm)}\left(
\sigma\right)  \right\vert \left\vert \Phi\right\rangle \;=\;\frac{1}{2}\,\left(\,
\Psi^{(\pm)}\left(  z_{1}\right)  \otimes\Psi^{(\pm)}\left(  z_{2}^{\prime
}\right)  \,+\,e^{i\rho}\,\Psi^{(\pm)}\left(  z_{1}^{\prime}\right)  \otimes
\Psi^{(\pm)}\left(  z_{2}\right) \, \right)
\]

Consequently, the full projected entangled wave function takes the form:%

\begin{gather}
\left\langle \Psi\left(  \omega\right)  \Psi\left(  \sigma\right)  \right\vert
\left\vert \Phi\right\rangle =\frac{1}{2}\left(  1-\left\vert \omega
\right\vert ^{2}\right)  ^{1/4}\left(  1-\left\vert \sigma\right\vert
^{2}\right)  ^{1/4}\underset{n,m=1,2...}{%
{\displaystyle\sum}
}\frac{\left(  \omega^{\ast}/2\right)  ^{2n}}{\sqrt{2n!}}\frac{\left(
\sigma^{\ast}/2\right)  ^{2m}}{\sqrt{2m!}}e^{-2\,\left(  n^{2}+m^{2}\right)
}\times\\
\times\left[  e^{2\left(  \left(  l-i\varphi\right)  n+\left(  l^{\prime
}-i\varphi^{\prime}\right)  m\right)  }\left(  1+\left(  1-\left\vert
\omega\right\vert ^{2}\right)  ^{1/2}\,\frac{\left(  \omega^{\ast
}\,e^{\,\left(  l-i\varphi\right)  }/2\right)  }{\sqrt{2n+1}}\;e^{-(2n+1/2)}%
\,\right)  \right. \nonumber\\
\left(  1+\left(  1-\left\vert \sigma\right\vert ^{2}\right)  ^{1/2}%
\,\frac{\left(  \sigma^{\ast}\,e^{\,\left(  l^{\prime}-i\varphi^{\prime
}\right)  }/2\right)  }{\sqrt{2m+1}}\;e^{-(2m+1/2)}\,\right) \\
+e^{i\rho}e^{2\left(  \left(  l-i\varphi\right)  m+\left(  l^{\prime}%
-i\varphi^{\prime}\right)  n\right)  }\left(  1+\left(  1-\left\vert
\omega\right\vert ^{2}\right)  ^{1/2}\,\frac{\left(  \omega^{\ast
}\,e^{\,\left(  l^{\prime}-i\varphi^{\prime}\right)  }/2\right)  }{\sqrt
{2n+1}}\;e^{-(2n+1/2)}\,\right) \nonumber\\
\left.  \left(  1+\left(  1-\left\vert \sigma\right\vert ^{2}\right)
^{1/2}\,\frac{\left(  \sigma^{\ast}\,e^{\,\left(  l-i\varphi\right)
}/2\right)  }{\sqrt{2m+1}}\;e^{-(2m+1/2)}\,\right)  \right]
\end{gather}

\subsection{Entanglement and Minimal Group Representation in the Cylinder}

In this case, we will analyze the unique and fundamental projection due to the
underlying metaplectic structure for each of the particular sectors of the full
Hilbert space. The wave function and the probability (the square norm in the
strict sense) will be consequently studied. 
\begin{itemize}
\item {In this case we must emphasize
the importance of the exponential factors that behave as $\varpropto $ $%
e^{-n^{2}}$ which are suppression factors for large $n$ and produce a rapid decay of the
analyzed functions, that is to say, the system truly classicalizes. }

\item{The other important point is the limit $\omega
\rightarrow \,\sigma $ which is the degeneracy of the analytical functions
describing the basic states of Mp(2): in the case of the entanglement
probability P$_{+-}$ for the even-odd sectors $(+ -)$ it is an unique and fundamental projection leaving only a vestige of the circular angular
variables as we will see explicitly in what follows:}
\end{itemize}

\subsection{Entanglement Probability P$_{+-}$ for the crossed Even-Odd (+-)
sectors}

In this case the Mp(n) projected wave function takes the form

\begin{gather}
\left\langle \Psi^{(+)}\left( \omega\right) \Psi^{(-)}\left( \sigma\right)
\right\vert \left\vert \Phi\right\rangle \;= \;\frac{1}{2}\left( 1-\left\vert
\omega\right\vert ^{2}\right) ^{1/4}\left( 1-\left\vert \sigma\right\vert
^{2}\right) ^{3/4}\underset{n,\,m\;=\;0,\,1,\,2...}{\sum }\frac{\left(
\omega^{\ast}/2\right) ^{2n}}{\sqrt{2n!}}\frac{\left( \sigma^{\ast}/2\right)
^{2m}}{\sqrt{2m!}}e^{-4\,\left( n^{2}+m^{2}\right) }\times \\
\times\left[\; e^{2\left( \left( l-i\varphi\right) n+\left( l^{\prime
}-i\varphi^{\prime}\right) m\right) }\,\frac{\left( \sigma^{\ast
}\,e^{\,\left( l^{\prime}-i\varphi^{\prime}\right) }/2\right) }{\sqrt {2m+1}}%
\;e^{-(2m+1/2)}\,\right. +\left. e^{i\rho}e^{2\left( \left(
l-i\varphi\right) m +\left( l^{\prime}-i\varphi^{\prime}\right) n\right) }\,%
\frac{\left( \sigma^{\ast}\,e^{\,\left( l-i\varphi\right) }/2 \right) }{%
\sqrt{2m+1}}\;e^{-(2m+1/2)}\;\right]  \notag
\end{gather}

\medskip

The square norm Probability P$_{+-}$

\begin{gather}
P_{+-}\;\equiv\;
\left\Vert \left\langle \Psi ^{(+)}\left( \omega \right) \Psi ^{(-)}\left(
\sigma \right) \right\vert \left\vert \Phi \right\rangle \right\Vert ^{2}=%
\frac{1}{2}\left( 1-\left\vert \omega \right\vert ^{2}\right) ^{1/2}\left(
1-\left\vert \sigma \right\vert ^{2}\right) ^{3/2} \\
\underset{n,\,m\,=\,0,1,2...}{{\displaystyle\sum }}\frac{\left( \left\vert
\omega \right\vert ^{2}/4\right) ^{2n}}{2n!}\frac{\left( \left\vert \sigma
\right\vert ^{2}/4\right) ^{2m+1}}{\left( 2m+1\right) !}e^{-4\,\left(
n^{2}+m^{2}\right) }\cdot  \notag \\
\cdot e^{-(4m+1)}\left[\; \left( e^{4\left( ln+l^{\prime }\left( m+1/2\right)
\right) } + e^{4\left( l^{\prime }n+l\left( m+1/2\right) \right) }\right)
\;+2\,e^{2\left( l+l^{\prime }\right) \left( n+m+1/2\right) }\cos \left(
2\,\left( \Delta \left( m-n\right) -\frac{\rho +1}{2}\right) \right) \right]
\notag
\end{gather}%
\newline
Let us notice the low dependence on the angular degeneracies (contained in $%
\Delta $) of the entanglement expressions for the different projections in
this case.

We consider now the limit  $\omega \rightarrow \,\sigma :$

\begin{gather}
\underset{\omega\rightarrow\,\sigma}{\lim}\left\langle \Psi^{(+)}\left(
\omega\right) \Psi^{(-)}\left( \sigma\right) \right\vert \left\vert
\Phi\right\rangle \;=\;\underset{\omega\rightarrow\,\sigma}{\lim}\left\{ \frac{1}{%
2}\left( 1-\left\vert \omega\right\vert ^{2}\right) ^{1/4}\left(1-\left\vert
\sigma\right\vert ^{2}\right) ^{3/4}\underset{n,m\,=\,0,1,2...}{{%
\displaystyle\sum}}\frac{\left(\omega^{\ast}/2\right)^{2n}}{\sqrt{2n!}}\frac{%
\left( \sigma^{\ast}/2\right)^{2m+1}}{\sqrt{2m+1!}}\times\right. \\
\left. \times e^{-4\,\left( n^{2}+m^{2}\right) } e^{-(2m+1/2)}\left[
e^{2\left( \left( l-i\varphi\right) n+\left(
l^{\prime}-i\varphi^{\prime}\right) m\right) }\,\,e^{\,\left(
l^{\prime}-i\varphi^{\prime}\right) }\;\,\right. +\left. e^{i\rho
}e^{2\left( \left( l-i\varphi\right) m+\left( l^{\prime}-i\varphi^{\prime
}\right) n\right) }\,\,e^{\,\left( l-i\varphi\right) }\;\,\right] \right\} 
\notag
\end{gather}

\bigskip The corresponding norm square Probability is\,:%
\begin{gather}
\underset{\omega \rightarrow \,\sigma }{\lim}\,\left\Vert \left\langle \Psi
^{(+)}\left( \omega \right) \Psi ^{(-)}\left( \sigma \right) \right\vert
\left\vert \Phi \right\rangle \right\Vert ^{2}= \\
=\underset{\omega \rightarrow \,\sigma }{\lim}\,\left\{ \frac{1}{2}\left(
1-\left\vert \omega \right\vert ^{2}\right) ^{1/2}\left( 1-\left\vert \sigma
\right\vert ^{2}\right) ^{3/2}\underset{n,\,m\;=\;0,\,1,\,2\,...}{{\displaystyle%
\sum }}\frac{\left( \left\vert \omega \right\vert ^{2}/4\right) ^{2n}}{2n!}%
\frac{\left( \left\vert \sigma \right\vert ^{2}/4\right) ^{2m+1}}{\left(
2m+1\right) !}\right. e^{-4\,\left( n^{2}+ m^{2}\right) }\cdot  \notag \\
\left. \cdot e^{-(4m+1)}\left[ \left( e^{4\left( ln+l^{\prime }\left(
m+1/2\right) \right) }+e^{4\left( l^{\prime }n+l\left( m+1/2\right) \right)
}\right) \;+2\,e^{2\left( l+l^{\prime }\right) \left( n+m+1/2\right) }\cos
\left( 2\left( \Delta \left( m-n\right) -\frac{\rho +1}{2}\right) \right) %
\right] \right\}  \notag
\end{gather}%
\begin{gather*}
=\underset{n,\,m\;=\;0,\,1,\,2\,...}{{\displaystyle\sum }}\delta _{2n,2m+1}\text{ }%
e^{-4\,\left( n^{2}+m^{2}\right) }\cdot \\
e^{-(4m+1)}\left[ \left( e^{4\left( ln+l^{\prime }\left( m+1/2\right)
\right) }+e^{4\left( l^{\prime }n+l\left( m+1/2\right) \right) }\right)
\;+2\;e^{2\left( l+l^{\prime }\right) \left( n+m+1/2\right) }\cos \left(
2\,\left( \Delta \left( m-n\right) -\frac{\rho +1}{2}\right) \right) \right]
\\
=\,\left[ \, 1+\cos \, \left( -\left(\, \Delta +\rho +1\,\right)\, \right)\, \right] 
\underset{n\;=\;0,\,1,\,2\,...}{{\displaystyle\sum }}e^{-8\,\left( n^{2}+3/4\right)
}\;2\;e^{4 \,\left(\, l^{\prime }\,+\,                   l\,\right)\, n}
\end{gather*}%
When $\left( l^{\prime }+l\right) = 0$ the entanglement Probability reduces to :%

\begin{equation*}
\underset{\omega \rightarrow \,\sigma }{\lim}\;\left\Vert \left\langle \Psi
^{(+)}\left( \omega \right) \Psi ^{(-)}\left( \sigma \right) \right\vert
\left\vert \Phi \right\rangle \right\Vert ^{2}\;=\;\left( \frac{1+\vartheta
_{3}\left( 0,e^{-8}\right) }{e^{6}}\right) \left( 1+\cos \left( \Delta +\rho
+1\right) \right)
\end{equation*}%

\bigskip

where $\vartheta _{3}\left( 0,e^{-8}\right) $ is the respective Theta
function. We can immediately see the dependence of the norm square on
the angle variables $(\varphi ,\varphi ^{\prime })$ through $\Delta = (\varphi -\varphi ^{\prime })$ in the
above limit.

\subsection{ Entanglement Probability P$_{++}$ for the Even (++) sectors}

In this case the wave function is

\begin{gather}
\left\langle \Psi ^{(+)}\left( \omega \right) \Psi ^{(+)}\left( \sigma
\right) \right\vert \left\vert \Phi \right\rangle =\frac{1}{2}\left(
1-\left\vert \omega \right\vert ^{2}\right) ^{1/4}\left( 1-\left\vert \sigma
\right\vert ^{2}\right) ^{1/4}\\
\underset{n,\,m\,=\,0,1,2...}{{\displaystyle\sum }}\frac{\left( \omega
^{\ast }/2\right) ^{2n}}{\sqrt{2n!}}\frac{\left( \sigma ^{\ast }/2\right)
^{2m}}{\sqrt{2m!}}e^{-2\,\left( n^{2}+m^{2}\right) }\left[ e^{2\left( l\text{%
\textit{n}+}l^{\prime }m-i\left( \varphi n+\varphi ^{\prime }m\right)
\right) }+e^{i\rho }e^{2\left( l\text{\textit{n}+}l^{\prime }m-i\left(
\varphi n+\varphi ^{\prime }m\right) \right) }\right]  \notag
\end{gather}%
And the norme square Probability is:%

\begin{gather}
\left\Vert \left\langle \Psi ^{(+)}\left( \omega \right) \Psi ^{(+)}\left(
\sigma \right) \right\vert \left\vert \Phi \right\rangle \right\Vert ^{2}\;=\;%
\frac{1}{2}\left( 1 - \left\vert \omega \right\vert ^{2}\right) ^{1/2}\left(
1 - \left\vert \sigma \right\vert ^{2}\right) ^{1/2}\\
\underset{n,\,m\;=\;0,\,1,\,2\,...}{{\displaystyle\sum }}\frac{\left( \left\vert
\omega \right\vert ^{2}/4\right) ^{2n}}{2n!}\frac{\left( \left\vert \sigma
\right\vert ^{2}/4\right) ^{2m}}{2m!}\,e^{\,-4\,\left( n^{2}+m^{2}\right) }\, \times\\
\times \,\left[\;
e^{4\,\left(\, l\text{\textit{n}+}l^{\prime }m\,\right) }+e^{\,4\,\left(\, l\text{%
\textit{m}+}l^{\prime }\,n\,\right) } + 
2\;e^{\,2\left(\, l+l^{\prime }\,\right) \,\left(\,
m+n\,\right) }\cos\left(\, 2\,\Delta \,\left(\,m-n \,\right) + \rho \,\right)\;\right] 
\notag
\end{gather}

\medskip

We can see in all the cases the expressions for the entanglement Probabilities are similar.

\bigskip

For the limit $\omega \rightarrow \,\sigma :$%
\begin{gather}
\lim_{\omega \rightarrow \,\sigma }\left\Vert \left\langle \Psi ^{(+)}\left(
\omega \right) \Psi ^{(+)}\left( \sigma \right) \right\vert \left\vert \Phi
\right\rangle \right\Vert ^{2}=\frac{1}{2} \\
\underset{n,\,m\,=\,0,1,2...}{{\displaystyle\sum }}\delta
_{n,m}e^{-4\,\left( n^{2}+m^{2}\right) }\left[ e^{4\left( l\text{\textit{n}+}%
l^{\prime }m\right) }+ e^{\,4\,\left( \,l\text{\textit{m}+}l^{\prime }n\right)
}+2\,e^{2\,\left(\, l+l^{\prime }\,\right) \left(\, m+n\,\right) }\cos \,\left(\, 2\,\Delta\,
\left( m-n\right) +\rho \,\right) \,\right]  \notag \\
= \underset{n\;=\;0,\,1,\,2...}{{\displaystyle\sum }}e^{-8\,n^{2}+ 4\,n\,\left(\, l\text{%
+}l^{\prime }\,\right) }\left(\, 1+\cos \rho \,\right)
\end{gather}%

We can see immediately that there are\textbf{\ no dependence of the norm
square entanglement Probability on the angle variables }($\varphi ,\varphi ^{\prime }$) \textbf{\
through }$\Delta .$ Consequently, when $\left( l^{\prime }+l\right) =0$, then : %

\begin{equation*}
\underset{\omega \rightarrow \,\sigma }{lim}\;\left\Vert \left\langle \Psi
^{(+)}\left( \omega \right) \Psi ^{(+)}\left( \sigma \right) \right\vert
\left\vert \Phi \right\rangle \right\Vert ^{2}\;=\;\left[\; 1 + \vartheta _{3}\left(\,
0,e^{-8}\,\right) \,\right] \left( 1+\cos \rho \right)
\end{equation*}

\subsection{Entanglement Probability P$_{--}$ for the Odd (-\,-) sectors}

Finally, as before, the wave function is

\begin{gather}
\left\langle \Psi ^{(-)}\left( \omega \right) \Psi ^{(-)}\left( \sigma
\right) \right\vert \left\vert \Phi \right\rangle =\frac{1}{2}\left(
1-\left\vert \omega \right\vert ^{2}\right) ^{3/4}\left( 1-\left\vert \sigma
\right\vert ^{2}\right) ^{3/4} \\
\underset{n,\,m\,=\,0,1,2...}{{\displaystyle\sum }}\frac{\left( \omega
^{\ast }/2\right) ^{2n+1}}{\sqrt{2n+1!}}\frac{\left( \sigma ^{\ast
}/2\right) ^{2m+1}}{\sqrt{2m+1!}}e^{-4\,\left( n^{2}+m^{2}\right)
}e^{-2(m+n+1/2)}\,\times  \notag \\
\left[ e^{2\left( \left( l-i\varphi \right) (n+1)+\left( l^{\prime
}-i\varphi ^{\prime }\right) (m+1)\right) }\,\;\,\right. +\left. e^{i\rho
}e^{2\left( \left( l-i\varphi \right) (m+1)+\left( l^{\prime }-i\varphi
^{\prime }\right) (n+1)\right) }\;\,\right]  \notag
\end{gather}

\bigskip

And the respective norm square is:

\begin{gather}
\left\Vert \left\langle \Psi ^{(-)}\left( \omega \right) \Psi ^{(-)}\left(
\sigma \right) \right\vert \left\vert \Phi \right\rangle \right\Vert ^{2}=%
\frac{1}{2}\left( 1-\left\vert \omega \right\vert ^{2}\right) ^{3/2}\left(
1-\left\vert \sigma \right\vert ^{2}\right) ^{3/2} \\
\underset{n,m\,=\,0,1,2...}{{\displaystyle\sum }}\frac{\left( \left\vert
\omega \right\vert ^{2}/4\right) ^{2n+1}}{2n+1!}\frac{\left( \left\vert
\sigma \right\vert ^{2}/4\right) ^{2m+1}}{2m+1!}e^{-4\,\left(
n^{2}+m^{2}\right) }e^{-4(m+n+1/2)}\,\times  \notag \\
\times \left\{ \left[ e^{4\left( l(n+1/2)+l^{\prime }(m+1/2)\right)
}\,\;+\,e^{4\left( l^{\prime }(n+1/2)+l(m+1/2)\right) }\right]
\,+2\,e^{2\left( l+l^{\prime }\right) \left( m+n+1\right) }\cos \left(
2\,\left( \Delta (m-n)+\rho \right) \right) \right\}  \notag
\end{gather}%

The limit $\omega \rightarrow \,\sigma$ is :%
\begin{gather}
\underset{\omega \rightarrow \,\sigma }{lim}\;\left\Vert \left\langle \Psi
^{(-)}\left( \omega \right) \Psi ^{(-)}\left( \sigma \right) \right\vert
\left\vert \Phi \right\rangle \right\Vert ^{2}=\frac{1}{2}\underset{%
n,m\,=\,0,1,2...}{{\displaystyle\sum }}\delta _{m,n}e^{-4\,\left(
n^{2}+m^{2}\right) }e^{-4(m+n+1/2)}\,\times \\
\times \left\{ \left[ e^{4\left( l(n+1/2)+l^{\prime }(m+1/2)\right)
}\,\;+\,e^{4\left( l^{\prime }(n+1/2)+l(m+1/2)\right) }\right]
\,+2\,e^{2\left( l+l^{\prime }\right) \left( m+n+1\right) }\cos \left(
2\,\left( \Delta (m-n)+\rho \right) \right) \right\} =  \notag \\
=\underset{n\,=\,0,1,2...}{{\displaystyle\sum }}e^{-8\left(
n^{2}+n+1/4\right) }\,e^{4\left( l+l^{\prime }\right) (n+1/2)}\,\left[
1\,\;+\,\cos \left( 2\,\rho \right) \right]
\end{gather}%

\medskip

Again, as in the limit ${\omega \rightarrow \,\sigma }$ for the even (++) states the odd-odd (-\,-) Entanglament  
\textbf{is independent of the angular variables } $(\varphi ,\varphi ^{\prime })$\textbf{\
through }$\Delta .$ When $l+l^{\prime } = 0$\, we have :%
\begin{equation*}
\underset{\omega \rightarrow \,\sigma }{\lim}\,P_{--}\;=\;\frac{1}{2}\,\vartheta
_{2}\left( 0,e^{-8}\right) \,\,\left( \,1\,\;+\,\cos 2\rho \,\right)
\end{equation*}%

Let us notice the low dependence on the angular degeneracies (contained in
$\Delta$) of the entanglement expressions for the different projection classicalizations in
this case.

\section{Entanglement of the Coset Coherent states in the Circle}

\subsection{Coset Coherent States in the Circle}

For the sake of completeness, let us first summarize the steps to follow for the determination of the coset coherent states:\newline

{\bf (i)} The Coset G/H identification \ $\rightarrow E\left(\,  2\,\right)
/\mathbb{T}_{2}$, \thinspace\ being $\mathbb{T}_{2}$ the group of translations
$\{g_{x},g_{y}\}\in\mathbb{T}_{2}$.

{\bf(ii)} The Fiducial vector determination: It is annihilated by all the
generators $h$ of the stability subgroup $H$ and for instance, invariant under
the action of $H.$ We propose:\,
$ \left\vert \;A_{0}\;\right\rangle \;=\;A\left(  z,x,y\right)  \left\vert
\varphi\right\rangle $
where $\left\vert \varphi\right\rangle $ is the London (circle) state, that is
expanded in the $\left\vert n\right\rangle $ state of the harmonic oscillator
and 
\[
A\left(  z,x,y\right)  _{(\pm)}\;=\;\left(  e^{z}\;\pm\;e^{-z}\right)
x\;+\;\left(  \mp e^{z}\;+\;e^{-z}\right) y,
\]
such that we can see: \; $
\left(  e_{x}\;+\;e_{y}\right)  A\left(  z,x,y\right)  _{(\,\pm)}\;=\;0 $

\medskip

{\bf (iii)} The coherent state is defined as the action of an element of the coset
group on the fiducial vector $\left\vert A_{0}\right\rangle $, consequently
the coherent state (still unnormalized yet) takes the form:\newline%
\begin{align}
e^{-\alpha\,\partial_{\varphi}}\left\vert A_{0}\right\rangle  &  =  =\frac{1}{\sqrt{2\pi}}%
\mathcal{S}\left(  \alpha,\varphi\right)  \underset{\left\vert \varphi
-\alpha/2\right\rangle }{\underbrace{\underset{n = 0,1,2..}{\sum}e^{-i\left(
\varphi-\alpha/2\right)  n}\left\vert n\right\rangle }}\nonumber\\
&  = \frac{1}{\sqrt{2\pi}}\;\mathcal{S}\left(  \alpha,\varphi\right)
\;\left\vert \varphi-\alpha/2\;\right\rangle
\end{align}
 where $\alpha\in\mathbb{C}$ is an arbitrary complex parameter in the
element of the coset, which must be adjusted after normalization, being%
\[
\mathcal{S}\left(  \alpha,\varphi\right)  \; \equiv\;\left(  A_{+}\cos
\alpha+A_{-}\sin\alpha\right)
\]
This is  the
most general coherent state from the Klauder-Perelomov construction which normalization takes the form 
\begin{align*}
\left\langle \beta,\varphi^{\prime}\right\vert \left\vert \alpha
,\varphi\right\rangle \;
= \;\frac{1}{2\pi}\;\;\frac{\mathcal{S}\left(  \beta^{\ast},\varphi^{\prime
}\right)  \; \mathcal{S}\left(  \alpha,\varphi\right)  }{1-e^{-\,i\,\left(
\,\varphi-\varphi^{\prime}\,-\,\left(  \alpha\,-\,\beta^{\ast}\right)
/2\,\right)  }}%
\end{align*}
Then, the state is fully normalizable for $\varphi\rightarrow\varphi^{\prime}$
iff the parameter $\alpha$ have $\operatorname{Im}\alpha\;\neq\;0$ :%

\begin{equation}
\left\vert \,\left\vert \alpha,\varphi\right\rangle \,\right\vert ^{2}\;=
\;\frac{1}{2\pi}\;\;\frac{\mathcal{S}\left(  \alpha^{\ast},\varphi\right)  \;
\mathcal{S}\left(  \alpha,\varphi\right)  }{1-e^{-\,i\left(  \, \alpha^{\ast
}\,-\,\alpha\right)  /2}} \label{qn}%
\end{equation}
where:
\[
\mathcal{S}\left(  \alpha^{\ast},\varphi\right)  \mathcal{S}\left(
\alpha,\varphi\right)  =\left(  x^{2} + y^{2}\right)  \cosh\left(
2\operatorname{Im}\alpha\right)  -\left( x^{2}-y^{2}\right)  \sin2\left(
\operatorname{Re}\alpha-\varphi\right)  \,+\,2\,x\,y\cos2\left(  \operatorname{Re}%
\alpha-\varphi\right)
\]
Consequently, it solves the problem of the London states that are overcomplete
but clearly not normalizable when $\varphi\rightarrow\varphi^{\prime}$:
\begin{equation}
\label{qn2}\left\langle \varphi\right\vert \left\vert \varphi^{\prime
}\right\rangle \;=\;\frac{1}{2\pi}\underset{n\,=\,0,1,2..}{\sum}e^{i\left(
\varphi-\varphi^{\prime}\right)  n}\;=\;\frac{1}{2\pi}\frac{1}{1-e^{-i\left(
\varphi-\varphi^{\prime}\right)  }}%
\end{equation}

We see explicitely from these expressions Eq.(\ref{qn}), Eq.(\ref{qn2}) how
the general coherent states on the circle $\left\vert \alpha,\varphi
\right\rangle $ (with the coherent characteristic complex parameter $\alpha$)
solve the problem of the non normalizability of the known (London, 't Hooft)
$\left\vert \,\varphi\right\rangle $ states in the circle.

From Eq.(\ref{qn}) the normalized state coherent state $\left\langle
\varphi\right\vert \left\vert \varphi^{\prime}\right\rangle $ is:%

\begin{equation}
\left\vert \alpha,\varphi\right\rangle \;=\; \underset{\mathcal{N}%
}{\underbrace{\sqrt{1-e^{-\operatorname{Im}\alpha}}\;e^{i\arg\mathcal{S}}}%
}\underset{n\,=\,0,1,2..}{\sum}e^{-i\left(  \varphi-\alpha/2\right)  \,
n}\;\left\vert n\right\rangle \label{ccs}%
\end{equation}

\medskip

The identity resolved in a weak sense always for $\operatorname{Im}\alpha>0$\,%
clearly showing  the role played by the coherent state characteristic complex parameter $\alpha$. 

\subsection{Action of the Mp(2) Group on the Coset States in the Circle}

Again, let us look at the sector $s = 1/4$ of the Hilbert space spanned by the
$Mp(2)$ coherent states (unnormalized), the basic state is%
\[
\left\vert \Psi^{\left(  +\right)  }\left(  \omega\right)  \right\rangle
= \left(  1-\left\vert \omega\right\vert ^{2}\right)  ^{1/4}\underset
{n = 0,1,2..}{\sum}\frac{\left(  \omega/2\right)  ^{2n}}{\sqrt{2n!}}\,\left\vert
\, 2n\, \right\rangle
\]

On the other hand:%
\begin{equation}
\left\langle \alpha,\varphi\right\vert =\mathcal{N}^{\ast}\underset
{n = 0,1,2..}{\sum}e^{i\left(  \varphi-\alpha^{\ast}/2\right)  n}\,\left\langle
\,n \,\right\vert
\end{equation}
Therefore, we have%
\begin{align}
\left\langle \alpha,\varphi\right\vert \left\vert \Psi^{\left(  +\right)
}\left(  \omega\right)  \right\rangle 
&  =\frac{\left(  1-\left\vert
\omega\right\vert ^{2}\right)  ^{1/4}}{\sqrt{2\pi}}\underset{n\,=\,0,1,2..}{\sum
}\frac{\left(  \omega e^{i\left(  \varphi-\alpha^{\ast}/2\right)  }/2\right)
^{2n}}{\sqrt{\left(  2n\right)  !}}\nonumber\\
& \; = \;\frac{\left(  1-\left\vert \omega\right\vert ^{2}\right)  ^{1/4}}%
{\sqrt{2\pi}}\underset{n\,=\, 0,1,2..}{\sum}\frac{\left(  z^{\prime}/2\right)
^{2n}}{\sqrt{(2n)!}}%
\end{align}
where
\[
\omega\,e^{\,i\left(  \varphi\,-\,\alpha^{\ast}/2\,\right)  }\;=\;z\;
e^{\,-i\,\alpha^{\ast}/2\,}\;\equiv\; z^{\prime},%
\]
and we see that the analytic function in the disc is now modified by the complex phase
$(\,\varphi-\alpha^{\ast}/2\,)$.

Similarly, for the sector $s = 3/4$ of the $Mp(2)$ states, that is the odd (-) states, we have:%

\begin{align}
\left\langle \alpha,\varphi\right\vert \left\vert \Psi^{\left(  -\right)
}\left(  \omega\right)  \right\rangle  &  =\frac{\left(  1-\left\vert
\omega\right\vert ^{2}\right)  ^{3/4}}{\sqrt{2\pi}}\underset{n\,=\,0,1,2..}{\sum
}\frac{\left(  \omega e^{i\left(  \varphi-\alpha^{\ast}/2\right)  }/2\right)
^{2n+1}}{\sqrt{\left(  2n+1\right)  !}}\nonumber\\
&  =\frac{\left(  1-\left\vert \omega\right\vert ^{2}\right)  ^{3/4}}%
{\sqrt{2\pi}}\underset{n\,=\,0,1,2..}{\sum}\frac{\left(  z^{\prime}/2\right)
^{2n+1}}{\sqrt{\left( 2n+1\right)  !}}%
\end{align}

\bigskip

Notice that by taking the scalar product between the coset coherent state
$ \left\vert \alpha,\varphi\right\rangle  $ and the $Mp(2)$ coherent states 
$\left\vert \Psi^{\left(  -\right)  }\left(  \omega\right)  \right\rangle $, we
obtain two non-equivalent expansions in terms of analytical functions on the
disk for the sectors of the minimal representations $s = 1/4$ and  $ s =3/4$: {\bf even
and odd $n$ states} respectively, in the eigenstaes $\left\vert
\,n\,\right\rangle $ of the harmonic oscillator.  

\medskip

Consequently,
$\ \left(  \;\omega\,e^{\,i\,\left(  \,\varphi-\alpha^{\ast}/2\,\right)
}\;\equiv\;z^{\prime}\;\right)  $:
\begin{subequations}
\label{0}%
\begin{equation}%
\begin{array}
[c]{c}%
\\
\left\langle \,\alpha,\varphi\right\vert \left\vert \Psi^{(\pm)}\left(
z^{\prime}\right)  \,\right\rangle =\\
\end{array}
\left\{
\begin{array}
[c]{cc}%
\left(  1 - \left\vert z^{\prime}\right\vert ^{2}\right)  ^{1/4}\underset
{n\,=\,0,1,2..}{\sum}\frac{\left(  z^{\prime}/2\right)  ^{2n}}{\sqrt{2n!}} &
\text{\ \ (+): \ {\it even states}}\\
& \\
\left(  1 - \left\vert z^{\prime}\right\vert ^{2}\right)  ^{3/4}\underset
{n\,=\,0,1,2..}{\sum}\frac{\left(  z^{\prime}/2\right)  ^{2n+1}}{\sqrt{\left(
2n+1\right)  !}} & \text{\ \ (-):\ \ {\it odd states}}%
\end{array}
\right.
\end{equation}
\newline Therefore, for the total projected state, $\left\langle
\;\alpha,\varphi\,\right\vert \left\vert \;\Psi\left(  z^{\prime}\right)
\;\right\rangle $ :%

\begin{equation}
\label{0}\left\langle \;\alpha, \varphi\right\vert \left\vert \;\Psi\left(  z^{\prime
}\right)  \; \right\rangle \;= \;\left\langle \; \alpha, \varphi\right\vert \left\vert
\;\Psi^{(+)}\left(  z^{\prime}\right)  \; \right\rangle \; + \; \left\langle
\; \alpha, \varphi\right\vert \left\vert \;\Psi^{(-)}\left(  z^{\prime}\right)  \;
\right\rangle \;,
\end{equation}

We have%

\end{subequations}
\begin{equation}
\left\langle \,\alpha,\varphi\right\vert \left\vert \Psi\left(  z^{\prime
}\right)  \, \right\rangle =\left(  1-\left\vert z^{\prime}\right\vert
^{2}\right)  ^{1/4}\underset{n\,=\,0,1,2..}{\sum}\frac{\left(  z^{\prime
}/2\right)  ^{2n}}{\sqrt{2n!}}\left[  1+\left(  1-\left\vert z^{\prime
}\right\vert ^{2}\right)  ^{1/2}\frac{\left(  z^{\prime}/2\right)  }%
{\sqrt{2n+1}}\right]  \label{gg}%
\end{equation}
\newline Let us notice the following observations:

\medskip

\textbf{(i)} The analyticity condition of the function $\left\langle
\alpha, \varphi\right\vert \left\vert \Psi\left(  z^{\prime}\right)  \right\rangle $
on the disk now constrained taking into account $\left\vert \,z^{\prime
}\right\vert \,=\,\left\vert \omega\right\vert e^{-\operatorname{Im}\alpha
/2}\,<\,1$  occurs under the already accepted condition arising from
the normalization function.

\bigskip

\textbf{(ii)} The topology of the circle induced by the coset coherent state
$\left\vert\alpha,\varphi \right \rangle $ Eq. (\ref{ccs}) not only modifies
the phase of $\omega$ (e.g: $\omega\,e^{i\,\left(  \, \varphi-\alpha^{\ast}/2
\,\right)  } = z^{\prime}\,)$ but also the ratio of the disc due the
displacement generated by the action of the coset.

\bigskip

\textbf{(iii)} The square norm  of Eq.(\ref{gg}) is easily calculated giving
as a result the function:%
\begin{align*}
\left\vert \;\left\langle \alpha, \varphi\right\vert \left\vert \Psi\left(  z^{\prime
}\right)  \right\rangle \;\right\vert ^{2}  &  \;=\; \left(  1-\left\vert
\,z^{\prime}\right\vert ^{2}\right)  ^{1/2}\;\cosh \left(  \frac{\left\vert
z^{\prime}\right\vert ^{2}}{2} \right)  \;+\;\left(  1-\left\vert z^{\prime
}\right\vert ^{2}\right)  ^{3/2}\,\sinh\left(  \frac{\left\vert z^{\prime
}\right\vert ^{2}}{2}\right)  \;+\\
&  +\;\left(  1-\left\vert \,z^{\prime}\right\vert ^{2}\right)  ^{1/2}%
\;\operatorname{Re}\,(z^{\prime})\underset{n\;=\;0\,,1\,,2..}{\sum}\frac
{\left\vert z^{\prime}/2\right\vert ^{\,4n}}{2n\,!\,\left(  2n+1\right)
},\;\;\;
\end{align*} 
where $z^{\prime}\,=\,\omega\,e^{\,i\,\left(  \,\varphi
-\alpha^{\ast}/2\,\right)  }$.
It shows a decreasing tail as $n$ increases, and the analyticity,
in this case in the disc $\left\vert z^{\prime}\right\vert \;<\;1$, with the
same comments as in the items {\bf (i)-(ii)} above.

\bigskip

Les us now consider the {\bf Entanglement in this case}.  We define now
by analogy with the 
London states, the variables for the coset circle states in the Bargmann representation :%
\begin{align*}
z_{1}  &  \equiv\omega\,e^{\,i\,\left(  \,\varphi-\alpha^{\ast}/2\,\right)
}\;\;\;\rightarrow\;\;\left\langle \,\varphi,\alpha\right\vert \left\vert \Psi^{(\pm
)}\left(  \omega\right)  \,\right\rangle \equiv\Psi^{(\pm)}\left(
z_{1}\right),
& \\ 
\text{\ }z_{1}^{\prime}  &  \equiv\omega\,e^{\,i\,\left(  \,\varphi^{\prime
}-\alpha^{\prime\ast}/2\,\right)  }\;\rightarrow\;\;\left\langle \,\varphi^{\prime
},\alpha^{\prime}\right\vert \left\vert \Psi^{(\pm)}\left(  \omega\right)
\,\right\rangle \equiv\Psi^{(\pm)}\left(  z_{1}^{\prime}\right) 
& \\
z_{2}  &  \equiv\sigma e^{\,i\,\left(  \,\varphi-\alpha^{\ast}/2\,\right)
} \;\;\;\rightarrow\;\;\left\langle \,\varphi,\alpha\right\vert \left\vert \Psi^{(\pm
)}\left(  \sigma\right)  \,\right\rangle \equiv\Psi^{(\pm)}\left(
z_{2}\right)
& \\
\text{\ \ }z_{2}^{\prime}  &  \equiv\sigma e^{\,i\,\left(  \,\varphi^{\prime
}-\alpha^{\prime\ast}/2\,\right)  }\;\;\rightarrow\;\;\left\langle \,\varphi^{\prime
},\alpha^{\prime}\right\vert \left\vert \Psi^{(\pm)}\left(  \sigma\right)
\,\right\rangle \equiv\Psi^{(\pm)}\left(  z_{2}^{\prime}\right)
\end{align*}
Then, the {\bf Entanglement} expresses as:%
\[
\left\langle \Psi^{(\pm)}\left(  \omega\right)  \Psi^{(\pm)}\left(
\sigma\right)  \right\vert \left\vert \Phi\right\rangle =\frac{1}{2}\left(
\Psi^{(\pm)}\left(  z_{1}\right)  \otimes\Psi^{(\pm)}\left(  z_{2}^{\prime
}\right)  +e^{i\rho}\Psi^{(\pm)}\left(  z_{1}^{\prime}\right)  \otimes
\Psi^{(\pm)}\left(  z_{2}\right)  \right)
\]
Explicitly, for the completely projected states, namely,  \textbf{the full entangled wave
function}, we have:
\begin{gather} \nonumber
\left\langle \Psi\left(  \omega\right)  \Psi\left(  \sigma\right)  \right\vert
\left\vert \Phi\right\rangle \;= \;\frac{1}{2}\,Z_{1}^{\,1/4}\,Z_{2}^{\,\prime\,1/4}\underset{n,\,m\,=\,1,2...}{%
{\displaystyle\sum}
}\frac{\left(  z_{1}^{\ast}/2\right)^{2n}}{\sqrt{2n!}}\frac{\left(
z_{2}^{\prime\,\ast}/2\right)  ^{2m}}{\sqrt{2m!}}\times\\
\left(  1 + \,Z_{1}^{1/2}%
\frac{\left(  z_{1}^{\ast}/2\right)  }{\sqrt{2n+1}}\right)  \left(  1 + \,Z_{2}^{\,\prime\,1/2}\frac{\left(
z_{2}^{\prime\,\ast}/2\right)}{\sqrt{2m+1}}\right)  + 
 \;\frac{1}{2}\,Z_{1}^{\,\prime\,1/4}\,Z_{2}^{\,1/4}\underset
{n,\,m\,=\,1,2\,...}{%
{\displaystyle\sum}
}\frac{\left(  z_{1}^{\prime\,\ast}/2\right)^{2n}}{\sqrt{2n!}}\frac{\left(
z_{2}^{\ast}/2\right)^{2m}}{\sqrt{2m!}}\times\nonumber \\
  e^{i\rho}\left(  1 + Z_{1}^{\,\prime\,  1/2}\frac{\left(  z_{1}^{\prime\,\ast}/2\right)}{\sqrt{2n+1}%
}\right)  \left(  1 + Z_{2}
^{\,1/2}\frac{\left(  z_{2}^{\ast}/2\right)}{\sqrt{2m+1}}\right)  
\end{gather}

where the following convenient notation have been introduced:
\begin{equation} \label{Z}
Z_1 \equiv 1-\left\vert z_{1}\right\vert ^{2}, \;\;\; Z_1^{\prime}  \equiv 1-\left\vert z_{1}^{\,\prime}\right\vert ^{2}, \;\;\; Z_2 \equiv 1-\left\vert z_{2}\right\vert ^{2}, \;\;\;
Z_2^{\prime}  \equiv 1-\left\vert z_{2}^{\,\prime}\right\vert^{2}
\end{equation}

And the \textbf{full Entanglement Probability} takes the following form:
\begin{gather*}
\left\Vert \left\langle \Psi\left(  \omega\right)  \Psi\left(  \sigma\right)
\right\vert \left\vert \Phi\right\rangle \right\Vert ^{2} = \left \{  \frac{1}%
{2}\, Z_{1}^{\,1/4} \, Z_{2}^{\,\prime \, 1/4}\underset{n,\,m\,=\,1,2...}{%
{\displaystyle\sum}}\frac{\left(  z_{1}^{\ast}/2\right)^{2n}}{\sqrt{2n!}}\frac{\left(
z_{2}^{\prime\ast}/2\right)^{2m}}{\sqrt{2m!}}\,\times \right.\\
\left(1 + Z_{1} ^{1/2}%
\frac{\left( z_{1}^{\ast}/2\right)  }{\sqrt{2n+1}}\right)  \left(  1 + Z_{2}^{\,\prime \,1/2}\frac{\left(
z_{2}^{\prime\ast}/2\right)  }{\sqrt{2m+1}}\right) + \frac{1}{2} \, Z_{1}^{\,\prime \,1/4} Z_{2}^{\,1/4}\underset{n,\,m\,= \,1,2\,...}{%
{\displaystyle\sum}
}\frac{\left(  z_{1}^{\prime\ast}/2\right)^{2n}}{\sqrt{2n!}}\frac{\left(
z_{2}^{\ast}/2\right)^{2m}}{\sqrt{2m!}}\,\times\nonumber\\
   e^{i\rho}\left(1+ Z_{1}^{\,\prime
\,1/2}\frac{\left(  z_{1}^{\prime\ast}/2\right)
}{\sqrt{2n+1}}\right)  \left(  1 + Z_{2}^{\,1/2}\frac{\left(  z_{2}^{\ast}/2\right)  }{\sqrt{2m+1}}\right) \times \,
 \frac{1}{2}\,Z_{1}
^{1/4}\, Z_{2}^{\,\prime
\,1/4}\underset{n,\,m \,=\,1,2...}{%
{\displaystyle\sum}}\frac{\left( z_{1}/2\right)^{2n}}{\sqrt{2n!}}\frac{\left(z_{2}^{\prime
}/2\right)^{2m}}{\sqrt{2m!}}\;\times \\
\left( 1 + Z_{1}^{1/2}%
\frac{\left(  z_{1}/2\right)  }{\sqrt{2n+1}}\right)  \left(  1 + Z_{2}^{\prime \,1/2}\frac{\left(
z_{2}^{\prime}/2\right)  }{\sqrt{2m+1}}\right)\,  + \\
+ \; \frac{1}{2}\;Z_{1}^{\,\prime \,1/4}\, Z_{2}^{1/4}\underset{n,\, m\,=\,1,2...}{%
{\displaystyle\sum}
}\frac{\left(  z_{1}^{\prime}/2\right)^{2n}}{\sqrt{2n!}}\frac{\left(
z_{2}/2\right)^{2m}}{\sqrt{2m!}}\times 
\left. e^{-i\rho}\left(  1 + Z_{1}^{\,\prime \, 1/2}\frac{\left(  z_{1}^{\,\prime}/2\right)}%
{\sqrt{2n+1}}\right)  \left(  1 + Z_{2} ^{\,1/2}\frac{\left(  z_{2}/2\right) }{\sqrt{2m+1}}\right) \right \}
\end{gather*}
\\

\subsection{Entanglement and the Minimal Group Representation for the Coset States}

In this case, for the circle coset coherent states, the unique and fundamental projection due to the underlying
metaplectic structure is given by the same analytical functions $\omega
\equiv\sigma,$ $\ $ and $ (z_{1,\,}z_{1}^{\prime})\equiv (z_{2,\,}z_{2}^{\prime})$, being
the respective wave function :%
\begin{gather}
\left\langle \Psi^{(+)}\left(  \omega\right)  \Psi^{(-)}\left(  \omega\right)
\right\vert \left\vert \Phi\right\rangle \;=\; \frac{1}{2}\,
Z_{1}^{1/4}\, Z_{1}^{\prime \,3/4} \underset{n,\,m\,=\,1,2\,...}{%
{\displaystyle\sum}
}\frac{\left(  z_{1}^{\ast}/2\right)^{2n}}{\sqrt{\left( 2n\right) \, !}}%
\frac{\left( z_{1}^{\prime\,\ast}/2\right)^{2m+1}}{\sqrt{\left(2m+1\right)
\,!}}\; + \\
+ \;\frac{1}{2}\; Z_{1}^{\prime\,1/4}\, Z_{1}^{3/4}\; e^{i\rho}   \underset {n,\,m\,=\,1,2\,...}{%
{\displaystyle\sum}}\frac{\left( z_{1}^{\prime\,\ast}/2\right)^{2n}}{\sqrt{2n\,!}}\frac{\left(
z_{1}^{\ast}/2\right)^{2m+1}}{\sqrt{\left(2m+1\right)\,  !}}\nonumber
\end{gather}
And the corresponding square norm {\bf Entanglement Probability} is:%
\begin{gather} 
\left\Vert \left\langle \Psi^{(+)}\left(  \omega\right)  \Psi^{(-)}\left(
\omega\right)  \right\vert \left\vert \Phi\right\rangle \right\Vert ^{2} = \\ \nonumber
  Z_{1}^{\,\prime 
\,3/2}\, Z_{1}^{1/2}\,\cosh\left(
\left\vert z_{1}\right\vert ^{2}/4\right) \; \sinh\left(\left\vert
z_{2}^{\prime}\right\vert ^{2}/4\right)  + \, Z_{1}^{\,\prime \, ^{1/2}}\, Z_{1} ^{3/2}\cosh\left(  \left\vert z_{1}^{\prime}\right\vert
^{2}/4\right) \;
\sinh\left(  \left\vert z_{1}\right\vert ^{2}/4\right)
   + \nonumber \\ 
  +\;Z_{1}^{\prime}\,Z_{1}  \left[\,e^{-i\rho}\cosh\left(
z_{1}^{\ast}z_{1}^{\prime}/4\right)\;\sinh\left(  z_{1}^{\prime\ast}%
z_{1}/4\right)\; + \; e^{i\rho}\cosh\left(  z_{1}z_{1}^{\prime\ast}/4\right)\;
\sinh\left(z_{1}^{\prime}z_{1}^{\ast}/4\right) \, \right] \nonumber
\end{gather}
We can notice that the above expression  contains trigonometric factors because:
\[
z_{1}^{\ast}\,z_{1}^{\prime} = \left\vert \omega\right\vert^{2} e^{-i\,\Delta
}\,e^{i\,\left(\, \alpha-\alpha^{\prime\ast}\right)  },\text{
\ \ \ \ \ \ \ \ \ \ \ }z_{1}^{\ast^{\prime}}z_{1}=\left\vert \omega\right\vert
^{2}e^{i\Delta}\,e^{-i\left(  \alpha-\alpha^{\prime\ast}\right)  }\text{\ }%
\]
Therefore, e.g. $\cosh\left(  z_{1}^{\ast}z_{1}^{\prime}/4\right)
=\cosh\left(  \left\vert \omega\right\vert ^{2}e^{-i\Delta}e^{i\left(
\alpha-\alpha^{\prime\ast}\right)  }/4\right),$  it could be expanded
if one specifically knows the exponent $(\alpha-\alpha^{\prime\ast})$.

\bigskip

As in the previous cases, we can see now the particular projections for the wave
functions and the corresponding norm square probabilities:

\subsection{ Entanglement Probabilities}

{\bf - Entanglement  Probability P$_{++}$ of the Even (++) sectors:}
\begin{gather}
\left\Vert \left\langle \Psi^{(+)}\left(  \omega\right)  \Psi^{(+)}\left(
\sigma\right)  \right\vert \left\vert \Phi\right\rangle \right\Vert
^{2} = \left( Z_{1}
\,Z_{2}\right)^{1/2}/4 \\
\left[ \left(  \frac{Z_{1}^{\,\prime}}{Z_{1}} \right)^{1/2}\cosh\left(  \left\vert z_{1}^{\prime
}\right\vert ^{2}/4\right)  \cosh \left(  \left\vert z_{2}\right\vert
^{2}/4\right)    +               \left(  \frac{Z_{2}^{\,\prime}}{Z_{2}}\right)^{1/2}\cosh\left(  \left\vert z_{1}\right\vert ^{2}/4\right) \, \cosh\left(
\left\vert z_{2}^{\prime}\right\vert ^{2}/4\right) \, \right]\,  + \nonumber \\
  + \left(  
\frac{Z_{1}^{\,\prime}\, Z_{2}^{\,\prime}}{Z_{1}\, Z_{2}}\,\right)^{1/4}\left[\,  e^{-i\rho}\cosh\left(
z_{1}^{\ast}z_{1}^{\prime}/4\right) \, \cosh\left( z_{2}^{\prime\ast}%
z_{2}/4\right) + e^{i\rho}\cosh\left(  z_{1}z_{1}^{\prime\ast}/4\right)\,
\cosh \left(z_{2}^{\prime} z_{2}^{\ast}/4 \,\right)\,\right]
\end{gather}

{\bf - Entanglement Probability  P$_{+-}$ of the  Even-Odd (+-) sectors:}%

\begin{gather*}
\left\Vert \left\langle \Psi^{(+)}\left(  \omega\right)  \Psi^{(-)}\left(
\sigma\right)  \right\vert \left\vert \Phi\right\rangle \right\Vert
^{2} =  \left( Z_{1} ^{1/2}\,Z_{2}^{3/2}/4 \right)\\
\left\{  \left[\left(  \frac{Z_{1}^{\prime}}{ Z_{1}}\right)^{1/2}\cosh\left(  \left\vert
z_{1}^{\prime}\right\vert ^{2}/4\right)  \sinh\left(  \left\vert
z_{2}\right\vert ^{2}/4\right) 
+ \left(  \frac{Z_{2}^{\prime}}{ Z_{2}}\right)^{3/2}\cosh\left(  \left\vert z_{1}\right\vert ^{2}/4\right)  \sinh\left(
\left\vert z_{2}^{\prime}\right\vert ^{2}/4\right) \right]  \right. + \\
\left.  + \left(  \frac{Z_{1}^{\prime}}{ Z_{1}}\right)^{1/4}
  \left(  \frac{Z_{2}^{\prime}}{ Z_{2}}\right)^{3/4}      \left[  e^{-i\rho}\cosh\left(  z_{1}^{\ast}z_{1}^{\prime}/4\right)
\sinh\left(  z_{2}^{\prime\ast}z_{2}/4\right)  + e^{i\rho}\cosh\left(
z_{1}z_{1}^{\prime\ast}/4\right)\sinh\left(  z_{2}^{\prime}z_{2}^{\ast
}/4\right)\right]\right\}
\end{gather*}

\bigskip

\medskip

{\bf 
- Entanglement Probability P$_{--}$ of the Even-Even (--) sectors:}%

\begin{gather}
\left\Vert \left\langle \Psi^{(-)}\left(  \omega\right)  \Psi^{(-)}\left(
\sigma\right)  \right\vert \left\vert \Phi\right\rangle \right\Vert
^{2} =    \left(  Z_{1}\,Z_{2}^{\;3/2}/4\right) \\
\left\{  \left[ \left(  \frac{Z_{1}^{\prime}}{ Z_{1}}\right)^{3/2}\sinh\left(  \left\vert
z_{1}^{\prime}\right\vert ^{2}/4\right)  \sinh\left(  \left\vert
z_{2}\right\vert ^{2}/4\right)
 + \left(  \frac{Z_{2}^{\prime}}{ Z_{2}}\right)^{3/2}\sinh\left(  \left\vert z_{1}\right\vert ^{2}/4\right)  \sinh\left(
\left\vert z_{2}^{\prime}\right\vert ^{2}/4\right) \right]  \right.  +\\
\left.  +  \left(  \frac{Z_{1}^{\prime}\, Z_{2}^{\prime}}{Z_{1}\,Z_{2}}\right)^{3/4}  \left[ e^{-i\rho}\sinh\left(
z_{1}^{\ast}z_{1}^{\prime}/4\right)  \sinh\left(  z_{2}^{\prime\ast}%
z_{2}/4\right) + e^{i\rho}\sinh\left(  z_{1}z_{1}^{\prime\ast}/4\right)
\sinh\left(  z_{2}^{\prime}z_{2}^{\ast}/4\right)  \right]  \right\}
\end{gather}

Here we can see that being $ (\sigma ,\omega) $ and $ (\varphi ,\varphi ^{\prime})
$ involved in only one single variable $z$: $(z_{1},z_{2},z_{1}^{\prime},z_{2}^{\prime})$, the classicalization is
more evident and the limits can be computed as in the previous cases. Recall that $ Z:(Z_1, Z_2, Z_{1}^{\prime},Z_{2}^{\prime})$ just relates to ${\vert z\vert }^2$ Eq.(\ref{Z}):
$Z_i = ( 1 - {\vert z_i\vert }^2)$ for each $i= (1,2)$.   

\section{Comparisons of Entanglements: Schrodinger Cat States and Mp(2) States} 

{\bf States:} Let us first recall the correspondence between the Schrodinger cat
states and the basic (+) and (-) $Mp(2)$ states, namely $\left\vert 1/4\right\rangle $
and $\left\vert 3/4\right\rangle:$

\begin{equation}%
\begin{tabular}
[c]{|l|l|l|l|}\hline
$Heisenberg-Weyl$ \ \ \ \ \ \ \ \ \ \ \ \ \ \ \ \ \ \  &  & $Metaplectic$
$Mp(2)$ \ \ \ \ \ \ \ \ \ \ \ \ \ \ \ \ \ \ \ \ \  & \\\hline
$\ \ \ \ \ \ \ \ \ \ \ \ \ \ \ \ \ \left\vert \alpha_{+}\right\rangle $ &
$\longrightarrow$ & $\ \ \ \ \ \ \ \ \ \ \ \ \left\vert 1/4\right\rangle $ &
Even\\\hline
$\ \ \ \ \ \ \ \ \ \ \ \ \ \ \ \ \ \left\vert \alpha_{-}\right\rangle $ &
$\longrightarrow$ & $\ \ \ \ \ \ \ \ \ \ \ \ \left\vert 3/4\right\rangle $ &
Odd\\\hline
\end{tabular}
\ \ \ \ \label{t}%
\end{equation}

\bigskip

where explicitly the standard cat Schrodinger states are%

\[
\left\vert \alpha_{\pm}\right\rangle \;=\;\frac{1}{\sqrt{2\;\pm\;2\;e^{-2\left\vert
\alpha\right\vert ^{2}}}}\;\left[ \; \left\vert \;\alpha \;\right\rangle \;\pm\;\left\vert\,
-\alpha\,\right\rangle \,\right]
\;\]
\\
which simply describe standard coherent states (e.g., Heisenberg-Weyl), $\alpha$ being the typical complex displacement parameter.
From the density matrix point of view,
the $Mp(2)$ states are favored with respect to the cat states because of the principle of minimum representation: The even and odd {\bf Mp(2) states  are  irreducible representations while the even and odd cat states are not}. In 
the case of the standard cat states, the density matrices
for the even and odd sectors are:
\begin{equation}
\rho_{\pm\alpha}\;=\;\frac{1}{2\;\left(\;  1\;\pm \;e^{\;-2\;\left\vert \;\alpha\;\right\vert\;
^{2}}\;\right)  }\;\left[\; \left\vert\, \alpha\,\right\rangle \left\langle\,
\alpha\,\right\vert \,+\,\left\vert\, -\,\alpha\,\right\rangle \left\langle \,-\,\alpha\,
\right\vert \,\pm \,\left(\,  \left\vert\, -\,\alpha\,\right\rangle \left\langle
\,\alpha\,\right\vert \,+\,\left\vert \,\alpha\,\right\rangle \left\langle \,-\,\alpha \,
\right\vert \,\right) \, \right]  \label{+-}%
\end{equation}

while in the fundamental case of the minimal group representation $Mp(2)$, we have
a {\it diagonal representation} clearly differentiated
into the corresponding {\it even} and {\it odd} subspaces: %
\begin{equation}
\rho_{Mp\left(  2\right)  }\; = \;\left\{
\begin{array}
[c]{c}%
\left\vert \;1/4\;\right\rangle \left\langle \;1/4\;\right\vert \;\;\rightarrow \; \;even \; states\\
\left\vert \;3/4\;\right\rangle \left\langle \;3/4\;\,\right\vert \;\;\rightarrow \; \;odd \; states
\end{array}
\right.  \label{rmp}%
\end{equation}
\\
 The minimal group representation does also manifest in the fact that  for the cat states $\rho_{\pm\alpha}$ have a tail   while such tail does not appear in the fundamental expressions of $\rho_{Mp\left( 2\right)}$ for both $Mp\left( 2\right)$ even and odd sectors.
 The minimal Mp(2) representation {\bf is
 diagonal\,: classical and quantum descriptions are both represented}, and describe {\bf both continuum and discrete sectors}. (And this is particularly relevant in describing the fundamental {\it quantum substrate} of space time as we showed in our recent work Refs \cite {cirilo-Sanchez}, \cite {universe}, \cite{Symmetry-2024}. Moreover,  
the basic $Mp(2)$ even and odd states $\left\vert 1/4\right\rangle ,\left\vert 3/4\right\rangle$ \, do not require any extrinsic generation process as the cat states, and can do form a
generalized state of the type:
\[
\left\vert \Psi_{Mp(2)}\right\rangle _{gen}\;=\;\frac{1}{\sqrt{\left\vert
A\right\vert ^{2}\;\pm \;\left\vert B\;\right\vert ^{2}}}\;\left[\;  A\;\left\vert\,
1/4\,\right\rangle \;\pm \;B\;\left\vert \,3/4\;\right\rangle \;\right]
\]
\\
(similar to the standard Schrodinger cat state $\left\vert \alpha_{\pm}\right\rangle $), and giving in
this case the following density matrix:
\begin{equation}
\rho_{Mp(2)_{gen}}=\frac{1}{\left\vert A\right\vert ^{2}\pm\left\vert
B\right\vert ^{2}}\left[\,  \left\vert A\right\vert ^{2}\;\left\vert
1/4\right\rangle \left\langle 1/4\right\vert \,+\,\left\vert B\right\vert
^{2}\;\left\vert 3/4\right\rangle \left\langle 3/4\right\vert 
\,\pm\,\left(
B^{\ast}A\,\left\vert 3/4\right\rangle \left\langle 1/4\right\vert +A^{\ast
}B\left\vert 1/4\right\rangle \left\langle 3/4\right\vert\right)\,\right]
\end{equation}
with coefficients A and B which can be fully computed 
depending on the  problem considered.

\bigskip

{\bf Entanglements:}  With respect to the entanglement states in Ref \cite{Symmetry-2024},
Schrodinger cat states are %
\begin{equation*}
\left\vert \alpha _{+}\right\rangle = e^{-\frac{1}{2}\,\left\vert \alpha
\right\vert ^{2}}\;\underset{n=0}{\overset{\infty }{\sum }}\;\frac{\alpha ^{2n}}{%
\sqrt{\left( 2n\right) !}}\;\left\vert 2n\right\rangle
\end{equation*}%
\begin{equation*}
\left\vert \alpha -\right\rangle = e^{-\frac{1}{2}\left\vert \alpha
\right\vert ^{2}}\;\underset{n=0}{\overset{\infty }{\sum }}\;\frac{\alpha ^{2n+1}%
}{\sqrt{\left( 2n+1\right) !}}\;\left\vert 2n+1\right\rangle
\end{equation*}%
\\
where
the standard displacement operators $\mathcal{D}\left( \alpha \right) $ and $%
\mathcal{D}\left( -\alpha \right) $ have been combined  operating on the
fiducial of the harmonic oscillator, namely $\left\vert 0\right\rangle $, to
sweep out {\it even and odd} $n$-states. The {\bf cat representation is not irreducible}, as
is the case of the $Mp(n)$ basic states, which do span out irreducible minimal subspaces.

The wave functions (projections) of the Schrodinger cat states 
on the London states are:%
\begin{equation*}
\left\langle \varphi \right\vert \left\vert \alpha _{+}\right\rangle \,= \,\frac{%
e^{-\frac{1}{2}\left\vert \widetilde{\alpha }\right\vert ^{2}}}{2\pi }%
\,\underset{n=0}{\overset{\infty }{\sum }}\,\frac{\widetilde{\alpha }^{2n}}{%
\sqrt{\left( 2n\right) !}}
\end{equation*}%
\begin{equation*}
\left\langle \varphi \right\vert \left\vert \alpha _{-}\right\rangle \,= \, \frac{%
e^{-\frac{1}{2}\left\vert \widetilde{\alpha }\right\vert ^{2}}}{2\pi }%
\,\underset{n=0}{\overset{\infty }{\sum }}\,\frac{\widetilde{\alpha }^{2n+1}}{%
\sqrt{\left( 2n+1\right) !}}
\end{equation*}%
where $\widetilde{\alpha }=\alpha\, e^{i\varphi }.$ 
Therefore, the total or
complete projected cat state on the circle is given by: 
\begin{equation}
\;\left\langle \varphi \right\vert \left\vert \alpha _{+}\right\rangle
\;+\;\left\langle \varphi \right\vert \left\vert \alpha _{-}\right\rangle \;= \;
\frac{e^{-\frac{1}{2}\left\vert \alpha \right\vert ^{2}}}{2\pi }\;\underset{n=0%
}{\overset{\infty }{\sum }}\;\frac{\widetilde{\alpha }^{n}}{\sqrt{n!}}
\end{equation} 
\\
which is just the projected London state $\varphi$ with the  Heisenberg-Weyl standard coherent
state  $\left\vert\, \alpha \,\right\rangle $ and the complex parameter $\alpha $. {\bf This contrasts with the case of projecting the $Mp(2)$ states}  where the complete
projected state on the circle is given by:

\begin{equation}
\left\langle \;\varphi \;\right\vert \left\vert \;\Psi \left( \omega \right)
\;\right\rangle \;=\;\frac{\left( 1-\left\vert z\right\vert ^{2}\right)
^{1/4}}{\sqrt{2\pi }}\underset{n\,=\,0,1,2..}{\sum }\frac{\left( z/2\right)
^{2n}}{\sqrt{(2n)\,!}}\left[ \,1\;+\;\left( 1-\left\vert z\right\vert
^{2}\right) ^{1/2}\frac{\left( z/2\right) }{\sqrt{2n+1}}\;\right]
\end{equation}%

where we see the weight function $\left( 1-\left\vert z\right\vert ^{2}\right)
^{1/4}$ indicating 
the geometry of the minimum irreducible subspace (in the even (+)  treated here) and  
 $0\leq \left\vert z\right\vert <1$ (the unitary disc). Recall too that the 
cat even and odd $ n$ representations are {\bf not} irreducible.

\begin{figure}
[ptb]
\begin{center}
\includegraphics[
height=2.4742in,
width=6.9029in,
]
{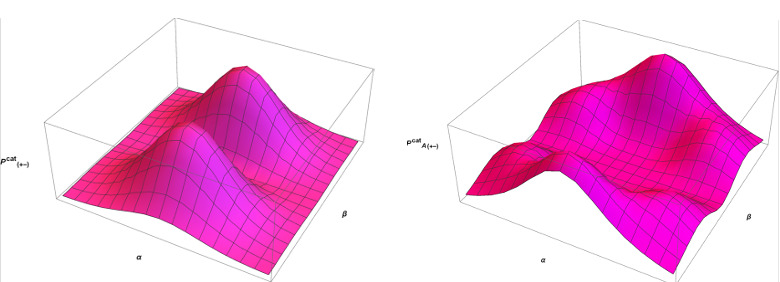}
\caption{ {\bf Schrodinger
cat and circle Entanglement Probability of even and odd states: $P^{\,cat} {(+- )}$ (left side) and the Antipodal one $P_{A} 
^{\,cat}{(+- )}$ (right side)} {\bf of the circle entangled orthogonal states}  $\varphi = \varphi^{\prime} + \pi/2 : \Delta \rightarrow \pi /2$ {projected onto the Schrodinger
cat even and odd states}. The
parameters $\protect\alpha $ and $\protect\beta $ in the Figure represent
the respective norms,  being the range of the parameters \, $ 0\leq \left\vert 
\protect\alpha \right\vert$ and $\left\vert \protect\beta \right\vert \,< 2$ to
appreciate the shape of the Entanglement Probability.}%
\label{f3}%
\end{center}
\end{figure}

\begin{figure}
[ptb]
\begin{center}
\includegraphics[
height=2.4742in,
width=6.9029in,
]
{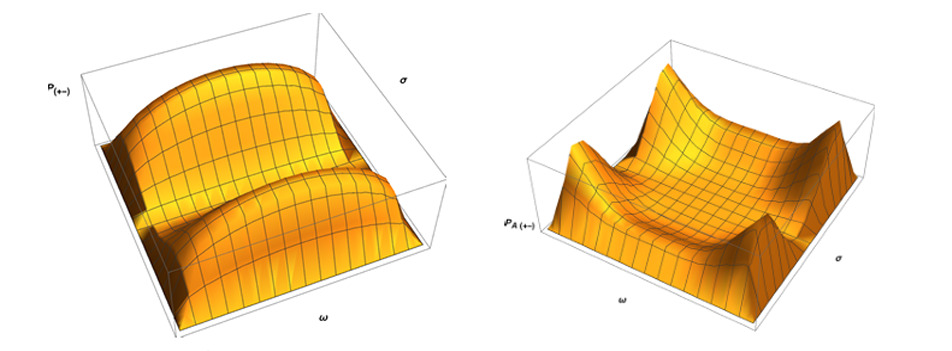}
\caption{{\bf Mp(2) and circle Entanglement Probability of even and odd states: $P{(+-)}$ (left side)\ and the Antipodal one $
P_{A}(+ -)$ (right side)} of the circle orthogonal states $ \varphi = \varphi^{\prime} + \pi/2 , ( \Delta \rightarrow \protect\pi /2 )$, {\bf projected on the
basic Mp(2) states:  Classicalization}. The variables $
\protect\omega $ and $\protect\sigma $ in the Figure representing the
respective norms of the analytic functions in the unitary disk, have the
range \,$0 \,\leq \left\vert \protect\omega \right\vert $
and $\,\left\vert \protect\sigma \right\vert \,< 1$.}%
\label{f4}%
\end{center}
\end{figure}

\begin{itemize}
\item{ {\bf In Figure \ref{f3}} we show
the Entanglement Probability $P_{ + \,- }^{cat}$ for the circle (London) states $\varphi$ projected via
the Schrodinger cat states for the orthogonal states $\varphi = \varphi^{\prime} + \pi/2 : \Delta \rightarrow \pi /2$. 
{\bf In Figure \ref{f4}}  we reproduce the Entanglement Probability $P_{ + \,- }$ for these circle (London) states $\varphi$ but projected via the $Mp(2)$ states:
Irreducible sectors of the minimal Hilbert space yielding {\bf Classicalization}.
 By comparing both cases we can see the geometrical difference between the
Entanglement probabilities:  $P_{+-}^{\,cat}$ projecting  with the Schrodinger cat
states versus $P_{+-}$  projecting with the basic $Mp(2)$ states.} 

\item{{\bf Cat Antipodal Entanglements:  Figure \ref{f3}} shows $P_{+-}^{\,cat}$ (left side)\ and the {\bf Antipodal Entanglement $P_{A + -}^{\,cat}$} of the  circle London states for $\Delta \rightarrow \protect\pi /2$, ie orthogonal states $\varphi = \varphi^{\prime} + \pi/2$.  The
variables $\protect\alpha $ and $\protect\beta $ in the Figure  are in the range \,$0\,\leq \left\vert 
\protect\alpha \right\vert ,\;\left\vert \protect\beta \right\vert \,< 2$ to
appreciate the shape of the Entanglement probability.}
\item{ {\bf Mp(2) Antipodal   Entanglements:} $P_{+-}$ \,(left side) and the {\bf Antipodal Entanglement} $%
P_{A + - }$ of the circle London orthogonal states  $\Delta \rightarrow \protect\pi /2$ are shown in {\bf Figure 4}. The variables \, $\protect\omega $ and $\protect\sigma $  of the  analytic functions in the unitary disk  (which square norms are displayed in the Figure) being in the range : $0\;\leq \left\vert \protect\omega \right\vert
,\;\,\left\vert \sigma \right\vert \,< \,1$.}
\end{itemize}


\section{Concluding Remarks}

In this paper we have computed and analyzed the Entanglement  of quantum physics within the new framework of our recent  work Ref \cite%
{APL25} on Classicalization. Thus, the results of our paper here are twofold :  {\bf Entanglement and Classicalization and the relationship between them}. 
In Section I we summarized the  main results of the study performed here and we do not summarize them here again. In this Section we just briefly highlight main concluding remarks in a synthetic way. We consider various types of  states on the circle : London states,  new coherent coset states and states with cylinder topology as eigenstates of a lowering operator. 

\begin{center}
{\bf We highlight here the following conclusions of this paper:}
\end{center}

{\bf(1) The Entanglement study} performed here clearly supports the fact the metaplectic group naturally classicalizes any quantum state as the realization of the  Minimal Group Representation Principle: this is uniquely realized by the Metaplectic group with its associated minimal Hilbert Space sectors, and the Entanglements of the different states show too {\bf Minimal Entanglements, in particular zero} when projecting on the Mp(2) states: {\bf Classicalization}).

\bigskip

{\bf(2) The result} of the item (1)  provides a precise conceptual and computational support  to the meaning of classicalization with respect to the condition of taking a limit for 
$\hslash $ (or any other control parameter) for the classical description which is more a formal limit rather than a full conceptual description for classicalization.

\bigskip

{\bf(3)} The projected Mp(2)  {\bf Entanglement Probability} is dependent on the angular variables, and through the difference $%
\Delta = (\varphi -\varphi^{\prime})$ in all cases treated here except for the
cylindrical coherent states. In the {\bf cylindrical topology case} the entanglement becomes {\bf independent of the angular variables} (observable) in the limit of coincident states $%
(\omega \rightarrow \sigma )$, projected on a single subspace (even or odd) of
the total Hilbert space. This is a {\bf quantum memory loss} manifestation in the Robertson sense, that is, there is no decoherence in the sense of interaction with an environment.

\bigskip

{\bf(4)  Entanglement Classicalization} is {\bf stronger} for the cylinder topology than for the circle:  Decay of the Entanglement Probability for large $n$ (even and odd levels  $n = 1, 2, 3, ...$) is decreasing exponentially with $(2n)^2$ and $(2n+1)^2$ in the cylinder, while it is decreasing exponentially with $(2n)$ and $(2n+1)$ in the circle.

\bigskip

{\bf(5) Comparison}  of the  Entanglements projected on the Mp(2) states and on other (not  Mp(2)) states clearly shows the differences between the two cases {\bf both for the Entanglement and Classicalization.  Figures \ref{f3} and \ref{f4}}  exhibit the differences between the Schrodinger cat projected Entanglement Probabilities  and the Mp(2) Entanglement Probabilities. 

\bigskip

{\bf(6)} {\bf Comparison} of 
the {\bf Antipodal Entanglement} (regulated by the phase control parameter $\rho = \pi$) with the Non Antipodal $(\rho = 0)$ one reveals interesting  properties in all cases studied here, for example  (as in {\bf Figure \ref{f4}})  the {\bf low and Minimal Entanglement} occurring in the Mp(2) projected states, as another manifestation of {\bf Classicalization}, particularly in the {\bf Antipodal Entanglement}.

\bigskip
 
 {\bf (7) Outlook and Implications:} The results of this paper on theoretical and conceptual aspects of quantum theory and its classicalization can have impact and applications in different branches of physics and other disciplines, classical and quantum information processing, quantum computation and experimental research, the quantum-classical interaction and interpretation measurements, the classical-quantum duality,  the classical or  quantum optimization. {\bf For instance: 
 
 \bf (i)} By choosing a type of geometry- topology of the states, one can obtain  stronger or lower classicalization. 
 
 {\bf (ii)} By using or avoiding coincident or orthogonal states, one can allows or avoid that entanglement be decreasing, broken or vanishing.  
 
 {\bf (iii)} By choosing Antipodal or Non Antipodal Entanglement (regulating the phase control parameter $\rho$ be equal to $\pi$ or $0$ respectively), Entanglement can be lower, minimal or not. 
 
 \medskip
 
{\bf (iv)} Combination of possibilities {\bf (i)-(ii)-(iii)} could yield more effects.

\section{Acknowledgements}

\medskip

D.J.C-L \, acknowledges  the Special Astrophysical Observatory of the Russian Academy of Sciences and CONICET of Argentine for Institutional and
financial support. \\
N.G.S \, acknowledges
useful communications with Gerard\, 't Hooft and
Jose Luis Mac Loughlin on various occasions.

\bigskip

\section{Appendix I\,: Mp(2) Projected Entanglement Probabilities in the Circle}

Below we provide now the explicit expressions of the square norm Entanglement Probabilities $P$ and their limits  $\omega
\rightarrow \sigma$ (analytic degeneracy) for the circle (London type) states.It is convenient to  introduce the following notation:
\begin{equation}
\beta = \frac{\vert \omega\vert^2}{4} \cos{\Delta}, \;\;\;\; \;\;  \widetilde{\beta } = \frac{\vert \omega\vert^2}{4} \sin{\Delta}
\end{equation}
\begin{equation}
\gamma = \frac{\vert \sigma\vert^2}{4} \cos{\Delta}, \;\;\;\; \;\;  \widetilde{\gamma } = \frac{\vert \sigma\vert^2}{4} \sin{\Delta}
\end{equation}

\bigskip

\textbf{(1) The Even-even sector Probability $P_{++}$} is given by:

\begin{gather*}
P_{++}\equiv \left\Vert \left\langle \Psi ^{(+)}\left( \omega \right) \Psi
^{(+)}\left( \sigma \right) \right\vert \left\vert \Phi \right\rangle
\right\Vert ^{2} = \frac{1}{2}\sqrt{\left( 1-\left\vert \omega \right\vert
^{2}\right) \left( 1-\left\vert \sigma \right\vert ^{2}\right) } \left\{
\cosh \beta \cosh \widetilde{\beta}
\;+\; \right. \\
\left. +\; \cos \rho \;  \left[\; \cosh \beta \cos \widetilde{\beta}  \cosh \gamma \cos \widetilde{\gamma} \;+ \; \sinh \beta \sin \widetilde{\beta} \sinh \sigma \sin \widetilde{\sigma} \;\right]\; + \right. \\
\left. + \;\sin \rho \;\left[ \sinh \beta \sin \widetilde\beta \cosh \gamma \cos \widetilde{\gamma}\; - \;\sinh \gamma \sin \widetilde{\gamma} \cosh \beta\cos \widetilde{\beta} \; \right]\, \right\}
\end{gather*}%
{\bf Limit $\omega \rightarrow \sigma :$}
\begin{gather*}
\underset{\omega \, \rightarrow \, \sigma }{\lim}\;P_{++}\;=\;\frac{1}{2}\left(
1-\left\vert \omega \right\vert ^{2}\right) \; \times \\
\left\{ \cosh ^{2}\left( \frac{\left\vert \omega \right\vert ^{2}}{4}\right)
+ \cos \rho \left[ \cosh ^{2}\left( \frac{\left\vert \omega \right\vert ^{2}}{4}\cos
\Delta \right) -\sin ^{2}\left( \frac{\left\vert \omega \right\vert ^{2}}{4}%
\sin \Delta \right) \right] \, \right\}
\end{gather*}

\bigskip

\bigskip

\textbf{(2) Even-odd sector Probability $P_{+-}$\,:} 

\begin{gather*}
P_{+-}\equiv \left\Vert \left\langle \Psi ^{(+)}\left( \omega \right) \Psi
^{(-)}\left( \sigma \right) \right\vert \left\vert \Phi \right\rangle
\right\Vert ^{2} = \frac{1}{2}\left( 1-\left\vert \omega \right\vert
^{2}\right)^{1/2}\left( 1-\left\vert \sigma \right\vert ^{2}\right)
^{3/2} \left\{ \cosh \left( \frac{\left\vert \omega \right\vert ^{2}}{4}%
\right) \sinh \left( \frac{\left\vert \sigma \right\vert ^{2}}{4}\right) + \right. \\
\left. + \left[\;\cosh \beta \cos \widetilde{\beta} \sinh \gamma\cos \widetilde{\gamma} \; + \;  \sinh \beta \sin \widetilde{\beta} \cosh \gamma \sin \widetilde{\gamma} \;\right] \cos \rho \;+ \right.\\
\left. + \left[ \;\cosh \beta \cos \widetilde{\beta}\cosh \gamma \sin \widetilde{\gamma} \;-\;\sinh \beta \sin \widetilde{\beta} \sinh \gamma \cos \widetilde{\gamma} \; \right] \sin \rho \;\right\}
\end{gather*}%

\bigskip

{\bf Limit $\omega \rightarrow \sigma :$}%
\begin{gather*}
\underset{\omega \:\rightarrow \;\sigma }{\lim }\;P_{+-}\; = \; \frac{1}{2}\left(
1-\left\vert \omega \right\vert ^{2}\right) ^{2}\left\{ \cosh \left( \frac{%
\left\vert \omega \right\vert ^{2}}{4}\right) \sinh \left( \frac{\left\vert
\omega \right\vert ^{2}}{4}\right) +\right. \\
\left. +\cosh \left( \frac{\left\vert \omega \right\vert ^{2}}{4}\cos \Delta
\right) \sinh \left( \frac{\left\vert \omega \right\vert ^{2}}{4}\cos \Delta
\right) \cos \rho +\cos \left( \frac{\left\vert \omega \right\vert ^{2}}{4}%
\sin \Delta \right) \sin \left( \frac{\left\vert \omega \right\vert ^{2}}{4}%
\sin \Delta \right) \sin \rho \right\}
\end{gather*}

\bigskip

\textbf{(3) Odd-odd sector Probability $P_{--}$\,:}

\begin{gather*}
P_{--}\equiv \left\Vert \left\langle \Psi ^{(-)}\left( \omega \right) \Psi
^{(-)}\left( \sigma \right) \right\vert \left\vert \Phi \right\rangle
\right\Vert ^{2} = \frac{1}{2}\left[ \left( 1-\left\vert \omega \right\vert
^{2}\right) \left( 1-\left\vert \sigma \right\vert ^{2}\right) \right]
^{3/2} \left\{\;\sinh \left( \frac{\left\vert \omega \right\vert ^{2}}{4}%
\right) \sinh \left( \frac{\left\vert \sigma \right\vert ^{2}}{4}\right)
+\right. \\
\left. +\left[ \;\sinh \beta \cos \widetilde{\beta} \sinh \gamma \cos \widetilde{\gamma} \;+\;\cosh \beta \sin \widetilde{\beta}  \cosh \gamma \sin \widetilde{\gamma}\;\right]\; \cos \rho \;\, + \right.\\
\left. +\;\left[\; \sinh \beta \cos \widetilde{\beta} \cosh \gamma \sin \widetilde{\gamma} \;-\;\cosh \beta  \sin \widetilde{\beta} \sinh \gamma \sin \widetilde{\gamma} \;\right]\; \sin \rho \;\right\}
\end{gather*}%

\bigskip

{\bf Limit $\omega \rightarrow \sigma :$}%
\begin{gather*}
\underset{\omega \;\rightarrow \;\sigma }{\lim }\;P_{--} \;= \;\frac{1}{2}\left(
1-\left\vert \omega \right\vert ^{2}\right) ^{3} \;\times \\
\left\{\; \sinh ^{2}\left( \frac{\left\vert \omega \right\vert ^{2}}{4}\right)
+\left[ \sinh ^{2}\left( \frac{\left\vert \omega \right\vert ^{2}}{4}\cos
\Delta \right) +\sin ^{2}\left( \frac{\left\vert \omega \right\vert ^{2}}{4}%
\sin \Delta \right) \right] \cos \rho \;\right\}
\end{gather*}%

\bigskip

As we see here, for the entangled Mp(2)-projected London states, the limits $%
\omega \rightarrow \sigma $ (analytic degeneracy) are $(\varphi ,\varphi
^{\prime })$ dependent through $\Delta \equiv (\varphi - \varphi
^{\prime }) $ in all cases.

\end{document}